\documentclass[fleqn, usenatbib]{mnras}
\usepackage{newtxtext, newtxmath} 	
\usepackage[T1]{fontenc}
\usepackage{graphicx}
\usepackage{amsmath}
\usepackage{mathtools}
\usepackage{multicol}
\usepackage{bm}
\usepackage{pdflscape}
\usepackage{natbib}
\usepackage[section]{placeins}
\usepackage{lipsum}
\usepackage{etoolbox}
\usepackage{tabularx}
\usepackage{xcolor}
\usepackage[toc, page]{appendix}
\usepackage{subfiles}
\hypersetup{ 
	colorlinks		= true,
	urlcolor 		= blue,
	linkcolor 		= blue,
	citecolor 		= blue
}

\providecommand{\noopsort}[1]{}

\newcommand{\Ctwelve}{\ensuremath{^{12}\text{C}}}
\newcommand{\Cthirteen}{\ensuremath{^{13}\text{C}}}
\newcommand{\Cfourteen}{\ensuremath{^{14}\text{C}}}
\newcommand{\Nthirteen}{\ensuremath{^{13}\text{N}}}
\newcommand{\Nfourteen}{\ensuremath{^{14}\text{N}}}
\newcommand{\Nfifteen}{\ensuremath{^{15}\text{N}}}
\newcommand{\Ofifteen}{\ensuremath{^{15}\text{O}}}
\newcommand{\Osixteen}{\ensuremath{^{16}\text{O}}}
\newcommand{\vice}{\texttt{VICE}}
\newcommand{\python}{\texttt{PYTHON}}
\newcommand{\nh}{\ensuremath{\text{[N/H]}}}
\newcommand{\feh}{\ensuremath{\text{[Fe/H]}}}
\newcommand{\oh}{\ensuremath{\text{[O/H]}}}
\newcommand{\no}{\ensuremath{\text{[N/O]}}}
\newcommand{\ofe}{\ensuremath{\text{[O/Fe]}}}
\newcommand{\ohno}{\no-\oh}
\newcommand{\msun}{\ensuremath{\text{M}_\odot}}
\newcommand{\persqkpc}{\ensuremath{\text{kpc}^{-2}}}
\newcommand{\refp}[1]{(\ref{#1})}
\newcommand{\rgal}{\ensuremath{R_\text{gal}}}
\newcommand{\hsim}{\texttt{h277}}
\newcommand{\subcc}{\ensuremath{_\text{cc}}}
\newcommand{\ycc}[1]{\ensuremath{y_\text{#1}^\text{CC}}}
\newcommand{\yia}[1]{\ensuremath{y_\text{#1}^\text{Ia}}}
\newcommand{\yagb}[1]{\ensuremath{y_\text{#1}^\text{AGB}}}

\defcitealias{Cristallo2011}{C11}
\defcitealias{Cristallo2015}{C15}
\defcitealias{Karakas2010}{K10}
\defcitealias{Karakas2016}{KL16}
\defcitealias{Karakas2018}{K18}
\defcitealias{Ventura2013}{V13}

\newcommand{\cristallo}{\citetalias{Cristallo2011}+\citetalias{Cristallo2015}}
\newcommand{\karakasten}{\citetalias{Karakas2010}}
\newcommand{\karakas}{\citetalias{Karakas2016}+\citetalias{Karakas2018}}
\newcommand{\ventura}{\citetalias{Ventura2013}}

\title[Empirical Constraints on the Nucleosynthesis of Nitrogen]{Empirical
Constraints on the Nucleosynthesis of Nitrogen}

\author[J.W. Johnson et al.]{James W. Johnson,$^{1, 2}$\thanks{
	Contanct e-mail: \href{mailto:
	johnson.7419@osu.edu}{johnson.7419@osu.edu}}
	David H. Weinberg,$^{1, 2, 3}$
	Fiorenzo Vincenzo,$^{1, 2, 4}$
	Jonathan C. Bird,$^{5}$ and
	\newauthor
	Emily J. Griffith$^{1, 2}$
	\\
	$^{1}$ Department of Astronomy, The Ohio State University,
	140 W. 18th Ave., Columbus, OH, 43210, USA
	\\
	$^{2}$ Center for Cosmology and Astroparticle Physics (CCAPP),
	The Ohio State University, 191 W. Woodruff Ave., Columbus, OH, 43210, USA
	\\
	$^{3}$ Institute for Advanced Study, 1 Einstein Dr., Princeton, NJ, 08540,
	USA
	\\
	$^{4}$ E.A. Milne Centre for Astrophysics, University of Hull, Cottingham
	Rd, Kingston upon Hull, HU6 7RX, United Kingdom
	\\
	$^{5}$ Department of Physics \& Astronomy, Vanderbilt University,
	2301 Vanderbilt Place, Nashville, TN, 37235, USA
}

\date{Accepted XXX; Received YYY; in original form ZZZ}
\pubyear{2021}

\begin{document}
\label{firstpage}
\pagerange{\pageref{firstpage}--\pageref{lastpage}}
\maketitle

\begin{abstract}
We derive empirical constraints on the nucleosynthetic yields of nitrogen by
incorporating N enrichment into our previously developed and empirically tuned
multi-zone galactic chemical evolution model.
We adopt a metallicity-independent (``primary'') N yield from massive stars and
a metallicity-dependent (``secondary'') N yield from AGB stars.
In our model, galactic radial zones do not evolve along the observed
[N/O]-[O/H] relation, but first increase in [O/H] at roughly constant [N/O],
then move upward in [N/O] via secondary N production.
By~$t\approx5$ Gyr, the model approaches an equilibrium [N/O]-[O/H] relation,
which traces the radial oxygen gradient.
We find good agreement with the [N/O]-[O/H] trend observed in extra-galactic
systems if we adopt an IMF-averaged massive star yield
$y_\text{N}^\text{CC}=3.6\times10^{-4}$, consistent with predictions for
rapidly rotating progenitors, and a fractional AGB yield that is linear in mass
and metallicity
$y_\text{N}^\text{AGB}=(9\times10^{-4})(M_*/M_\odot)(Z_*/Z_\odot)$.
This model reproduces the [N/O]-[O/H] relation found for Milky Way stars
in the APOGEE survey, and it reproduces (though imperfectly) the trends of
stellar [N/O] with age and [O/Fe].
The metallicity-dependent yield plays the dominant role in shaping the gas-phase
[N/O]-[O/H] relation, but the AGB time-delay is required to match the APOGEE
stellar age and [O/Fe] trends.
If we add~$\sim$40\% oscillations to the star formation rate, the model
reproduces the scatter in gas-phase [N/O] vs. [O/H] observed in external
galaxies by MaNGA.
We also construct models using published AGB yields and examine their empirical
successes and shortcomings.
For all AGB yields we consider, simple stellar populations release half their
N after only~$\sim$250 Myr.
\end{abstract}


\section{Introduction}
\label{sec:intro}

From a nucleosynthesis perspective, N is a unique element.
Along with C and He, it is one of only three elements lighter
than iron peak nuclei thought to owe a significant portion of its abundance
to asymptotic giant branch (AGB) stars~\citep[e.g.][]{Johnson2019}.
N is also the primary by-product of the CNO cycle, a cyclic nuclear reaction
that catalyses the conversion of H into He in stars more massive than the sun.
However, uncertainties surrounding the nucleosynthetic yields of N make it
difficult to model its abundances accurately.
Here we take an empirical approach to constrain N yields by using
state-of-the-art galactic chemical evolution (GCE) models to assess which
functional forms describing the yield can reproduce recent observational
data for gas phase abundances and trends in Milky Way disc stars found
by~\citet{Vincenzo2021}.

\begin{figure*}
\centering
\includegraphics[scale = 0.6]{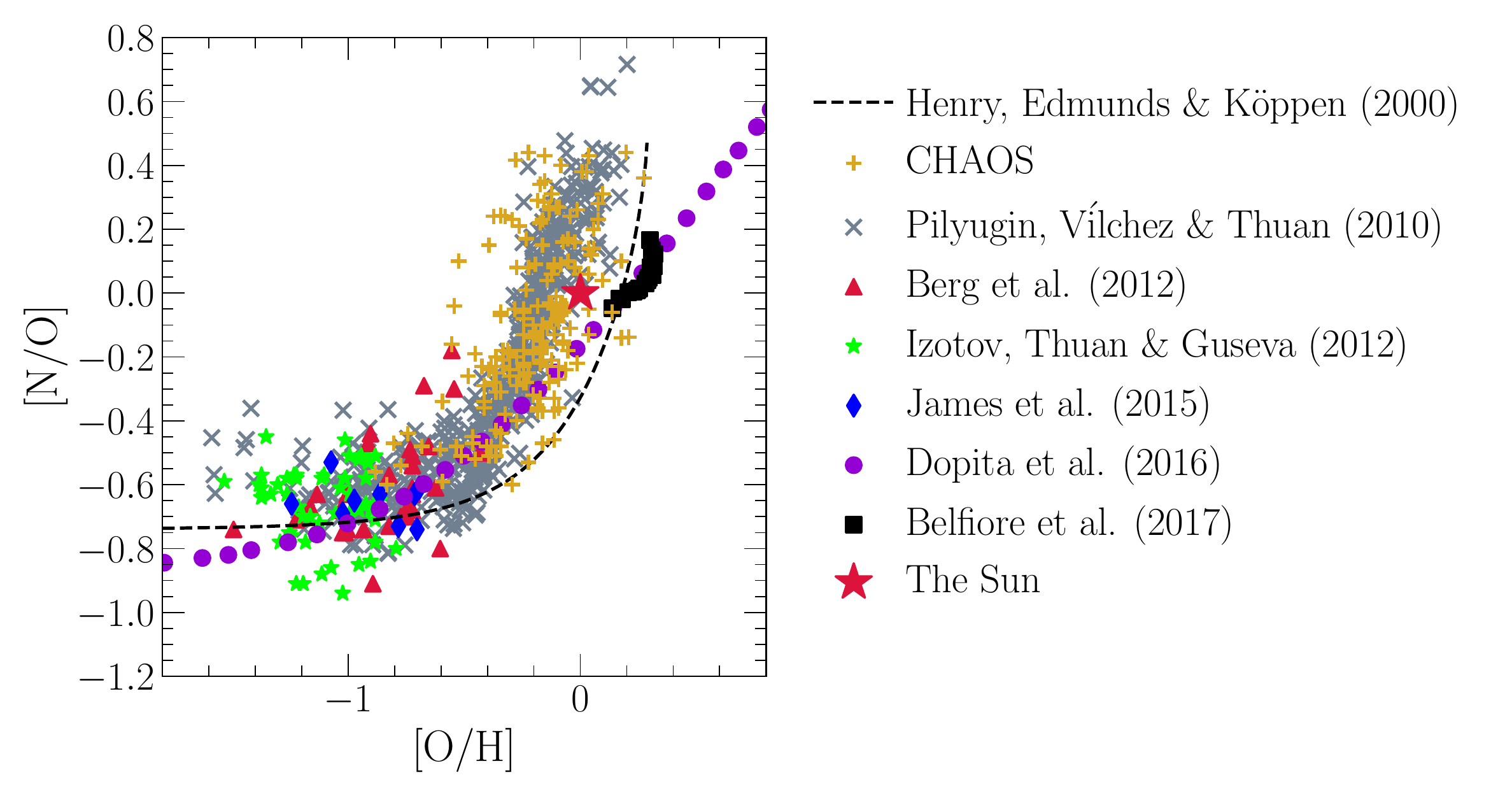}
\caption{
	The~\ohno~relation as observed in different galactic environments:
	HII regions from the first six CHAOS galaxies (golden +'s: NGC 3184, NGC
	628, NGC 5194, NGC 5457, M101, and NGC 2403;~\citealp{Berg2020,
	Skillman2020, Rogers2021}) and other nearby NGC spiral galaxies (grey X's;
	\citealp[][``ONS'' calibration]{Pilyugin2010}), HII regions in blue diffuse
	star forming dwarf galaxies (red triangles:~\citealp{Berg2012}; green stars:
	\citealp{Izotov2012}; blue diamonds:~\citealp{James2015}), in local stars
	and HII regions (purple circles:~\citealp{Dopita2016}), and in the MaNGA
	IFU survey (black squares:~\citealp{Belfiore2017}).
	The fit to~\no~as a function of~\oh~in Galactic and extragalactic HII
	regions by~\citet{Henry2000} is shown as a black dashed line.
	We omit all uncertainties for visual clarity.
	The Sun, at (0, 0) on this plot by definition, is marked by a large red
	star.
}
\label{fig:no_oh_observed}
\end{figure*}

Observationally, N abundances in external galaxies are generally measured in
the gas phase and are used as a metallicity indicator because of their strong
correlation with O abundances.
In Fig.~\ref{fig:no_oh_observed}, we present a compilation of such measurements
along with data from the Milky Way:
\begin{enumerate}
	\item[\textbf{1.}] HII regions in the first six CHAOS\footnote{
		CHAOS: CHemical Abundances Of Spirals~\citep{Berg2015}
	} galaxies: NGC 3184, NGC 628, NGC 5194, NGC 5457, M101, and NGC 2403
	\citep{Berg2020, Skillman2020, Rogers2021}.

	\item[\textbf{2.}] HII regions in nearby NGC spirals
	\citep*[][``ONS'' calibration]{Pilyugin2010}.

	\item[\textbf{3.}] HII regions in blue, diffuse star forming dwarf galaxies
	(\citealp{Berg2012};~\citealp*{Izotov2012};~\citealp{James2015}).

	\item[\textbf{4.}] Local stars and HII regions~\citep{Dopita2016}.

	\item[\textbf{5.}] Galactic and extragalactic HII regions
	\citep*{Henry2000}.

	\item[\textbf{6.}] Star-forming regions in 550 nearby galaxies in the
	MaNGA IFU\footnote{
		MaNGA: Mapping Nearby Galaxies at Apache Point Observatory
		\citep{Bundy2015}.
		IFU: Integral Field Unit.
	} survey~\citep{Belfiore2017}.
\end{enumerate}
Despite intrinsic scatter and some systematic variation in how the abundances
are determined, this~\ohno\footnote{
	We follow standard notation where [X/Y]
	$\equiv \log_{10}(X/Y) - \log_{10}(X/Y)_\odot$.
} relation is found to be similar across a wide range of astrophysical
environments.
Furthermore, recent arguments from both theoretical~\mbox{\citep{Vincenzo2018}}
and observational perspectives~\citep{HaydenPawson2021} suggest that this
relation is largely redshift-invariant.
Previous studies have interpreted this consistency as an indication that the
relation is nucleosynthetic in origin, reflective of a ``primary'' yield that
does not depend on a star's initial metal content and a ``secondary'' yield
that does (\citealp{VilaCostas1993};~\citealp*{vanZee1998};~\citealp{Henry1999,
PerezMontero2009};~\citealp*{Pilyugin2012};~\citealp{Andrews2013}).
Although we have highlighted star forming galaxies in
Fig.~\ref{fig:no_oh_observed}, N abundances are also easily measured in
massive ellipticals (see, e.g.,~\citealp{Schiavon2010},~\citealp{Conroy2013},
and~\citealp*{Conroy2014} for observational references), allowing it to
potentially bridge the gap between the physical processes affecting galaxies of
different morphologies.
\par
The challenge in interpreting N abundances is that accurate nucleosynthetic
yields from various enrichment channels remain elusive.
Relative to other light elements, N synthesis is difficult to model because it
is sensitive to uncertain details of stellar evolution, such as internal
mixiing (see discussion in, e.g.,~\citealp{Andrews2017} and
in~\S~\ref{sec:yields:ccsne} below).
In this paper, we constrain N yields empirically by testing the performance of
various assumptions within the framework of GCE models.
To this end we make use of the multi-zone model for the Milky Way published by
\citet{Johnson2021}, which treats the Galaxy as a series of concentric rings,
describing each one as a conventional one-zone model of chemical evolution
(see discussion in~\S~\ref{sec:multizone}).
This approach has been employed in the past to compute abundances for many
Galactic regions simultaneously (\citealp{Matteucci1989, Wyse1989, Prantzos1995,
Schoenrich2009};~\citealp*{Minchev2013, Minchev2014};~\citealp{Minchev2017};
\citealp*{Sharma2021}).
Because of the apparent universality of the~\ohno~relation, our results using
the Milky Way as a case test should apply to other galaxies as well.
\par
At low metallicity, rotating massive stars play a key role in establishing the
observed N abundances (\citealp*{Chiappini2003, Chiappini2005};
\citealp{Chiappini2006};~\citealp*{Kobayashi2011};~\citealp{Prantzos2018};
\citealp*{Grisoni2021}).
Rotation plays a pivotal role in stellar evolution, inducing effects such as
shear mixing, meridional circulation, and horizontal turbulence~\citep{Zahn1992, 
Maeder1998, Lagarde2012}.
These effects carry internally produced C and O nuclei into the H-burning shell
where they can be processed into~\Nfourteen~via the CNO cycle~\citep{Heger2010,
Frischknecht2016, Andrews2017}.
Metal-poor stars spin faster and are more compact~\citep*{Maeder1999}, making
these effects stronger and consequently enhancing N yields~\citep*{Meynet2002a,
Meynet2002b, Meynet2006}.
We find similar results here comparing various theoretical models for massive
star nucleosynthesis (see discussion in~\S~\ref{sec:yields:ccsne}).
\par
In sufficiently massive AGB stars, the base of the convective envelope is hot
enough to activate proton capture reactions, allowing the CNO cycle to convert
C and O isotopes into~\Nfourteen: a process known as hot bottom burning (HBB).
AGB stars are also known to experience thermal pulsations, and often these
pulsations are accompanied by a penetration of the convective enevelope into the
CO-rich core, which incorporates some of this material into the envelope
itself: a process known as third dredge-up (TDU).
When both processes are active, TDU adds new seed nuclei for HBB to turn
into~\Nfourteen, substantially increasing N yields.
We demonstrate in~\S\S~\ref{sec:yields:agb} and~\ref{sec:yields:imf_agb} that
various published theoretical models predict significantly discrepant N yields
for high mass AGB stars as a consequence of differences in TDU and
HBB.
The differences in these processes are in turn a consequence of the uncertain
microphysical assumptions built into stellar evolution models (e.g. mass
loss, opacity, convection and convective boundaries, nuclear reaction
networks).
In~\S~\ref{sec:results:yields}, we test the extent to which each of these
``off-the-shelf'' yield models are able to reproduce the~\ohno~relation in
GCE models.
\par
With a sample of 6,507 galaxies from the MaNGA IFU survey~\citep{Bundy2015},
\citet{Schaefer2020} demonstrate that the intrinsic scatter in
the~\ohno~relation at fixed galaxy mass is correlated with variations in the
local star formation efficiency (SFE).
In regions of slower star formation,~\no~tends to be slightly higher at
fixed~\oh~(see their fig. 4), which is expected from simple GCE models.
In classical ``closed-box models''~\citep[e.g.][]{Molla2006}, more AGB stars
enrich the interstellar medium (ISM) with N by the time a given~\oh~is reached,
whereas in ``open-box models'' with inflows and outflows like the ones we
present here, dilution by primordial gas accretion drives~\oh~down at fixed~\no.
However,~\citet{Schaefer2020} did not investigate stellar migration as a
potential source of additional scatter in the gas-phase~\ohno~relation.
In principle, there could be a deficit or surplus of N-producing AGB stars in a
given Galactic region at any time simply because the orbits are evolving,
driving additional scatter in the correlation.
The~\citet{Johnson2021} GCE model is an ideal tool with which to test this
hypothesis; the novel difference between theirs and previous models with
similar motivations is that it allows stellar populations to enrich
rings at different radii as they migrate.
Originally developed to study the abundances of O and Fe, this aspect of
Galactic evolution turned out to have an important impact on the delayed type
Ia supernova (SN Ia) enrichment of Fe, causing stochastic fluctuations in the
enrichment rates with time at fixed radius.
Here we use the same methodology to test for similar effects in the delayed AGB
star production of N, in turn assessing whether migration or variability in the
SFE dominate scatter in the~\ohno~relation.
\par
With stellar abundance data, we can test the N abundances predicted by our
model against observables unavailable for the gas phase, such as age and~\ofe.
Using data from the Apache Point Observatory Galaxy Evolution Experiment
(APOGEE;~\citealp{Majewski2017}) with asteroseismic mass measurements,
\citet{Vincenzo2021} demonstrate that when stellar N abundances are corrected
for internal mixing processes, the correlations with stellar age and other
elemental abundances are affected.
Whether or not our GCE model is able to reproduce their data constitutes a
valuable test of our understanding of N nucleosynthesis and the history of N
enrichment in the Milky Way.
\citet{Vincenzo2021} find good agreement between the APOGEE abundances and the
\citet{Dopita2016} data, which we find to be a good representation of external
galaxies as well; we therefore take the~\citet{Dopita2016} trend (the purple
points in Fig.~\ref{fig:no_oh_observed}) as our observational benchmark.
\par
In~\S~\ref{sec:yields}, we discuss our adopted yields of N from its dominant
nucleosynthetic sources.
We discuss the details of our multi-zone chemical evolution model
in~\S~\ref{sec:multizone}.
We describe the evolution of a fiducial model in~\S~\ref{sec:results:fiducial}.
In~\S~\ref{sec:results:yields}, we quantify the~\ohno~relation predicted by our
model with various ``off-the-shelf'' AGB star yield models taken from the
literature.
We investigate the relative importance of the delay-time distribution and the
metallicity-dependence of AGB star yields in~\S~\ref{sec:results:t_z_dep_comp}.
We compare our model predictions to stellar N abundances corrected for internal
mixing processes in~\S~\ref{sec:results:vincenzo_comp}.
We assess the sources of intrinsic scatter in the~\ohno~relation
in~\S~\ref{sec:results:schaefer_comp}.
We provide an analytic understanding of our key results
in~\S~\ref{sec:results:ohno_equilibrium} and summarize our conclusions
in~\S~\ref{sec:conclusions}.

\section{Nucleosynthesis}
\label{sec:yields}

\begin{figure*}
\centering
\includegraphics[scale = 0.65]{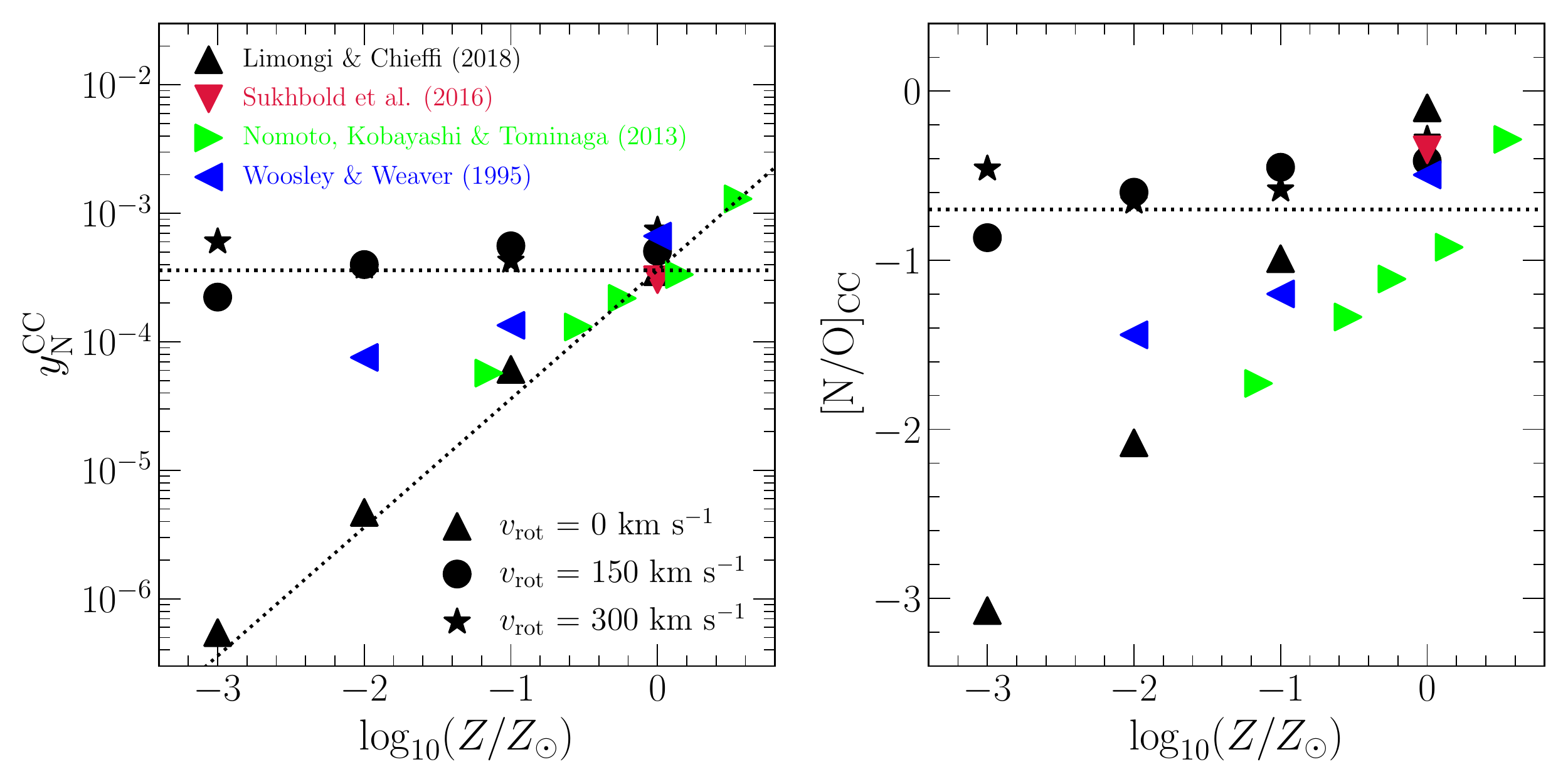}
\caption{
\textbf{Left}: IMF-averaged CCSN yields of N calculated using~\vice's
\texttt{vice.yields.ccsne.fractional} function with the tables published by
\citet[][blue]{Woosley1995},~\citet[][green]{Nomoto2013},
\citet[][red]{Sukhbold2016}, and~\citet[][black]{Limongi2018}.
All studies report yields for non-rotating progenitors, shown by the triangles;
for visual clarity, the triangles point in a different direction for each study
according to the legend.
\citet{Limongi2018} report additional yields for progenitors with rotational
velocities of 150 (circles) and 300 km/s (stars).
The horizontal dashed line marks~$\ycc{N} = 3.6\times10^{-4}$,
the value of our fiducial CCSN yield of N in our GCE models.
We also use the form shown by the slanted line (equation~\ref{eq:linear_yncc})
in~\S~\ref{sec:results:yields:yncc} in combination with some of our AGB star
yield models discussed in~\S~\ref{sec:yields:agb}.
\textbf{Right}: The~\no~ratio predicted by each of the explosion models in
the left-hand panel, under the same colour-coding and marker scheme.
We mark the position of~$\no = -0.7$ with a black dotted line, the value
roughly suggested by the observations of low-metallicity systems highlighted
in Fig.~\ref{fig:no_oh_observed}.
}
\label{fig:n_cc_yields}
\end{figure*}

Here we make use of the chemical evolution model for the Milky Way
presented by~\citet{Johnson2021}, which runs using the publicly available
\texttt{Versatile Integrator for Chemical Evolution}
(\vice, see Appendix~\ref{sec:vice};~\citealp{Johnson2020, Griffith2021a,
Johnson2021}), an open-source~\python~package designed for GCE modeling.
\citet{Johnson2021} focus their discussion of the model predictions on O and
Fe, and we retain their yields of these elements here.
The supernova (SN) yields are defined as the net mass of some element X
produced over all explosion events in units of the progenitor cluster's mass.
For example, with a yield of~$y_\text{X} = 0.001$, a hypothetical~$1000~\msun$
star cluster would produce~$1~\msun$ of the element X instantaneously in the
case of core-collapse supernovae (CCSNe) or over the delay-time distribution
(DTD) in the case of SNe Ia.
These yields are net yields in that they do not include the metal mass
ejected to the ISM that was initially present within a star; in the previous
example, the~$1~\msun$ yield is only the newly produced metal mass.
We adopt the following values from~\citet{Johnson2021}:
\begin{itemize}
	\item $\ycc{O} = 0.015$

	\item $\ycc{Fe} = 0.0012$

	\item $\yia{O} = 0$

	\item $\yia{Fe} = 0.00214$,
\end{itemize}
where the subscripts and superscripts differentiate between the element and the
SN type.
These choices are based on a mix of theoretical and empirical considerations.
For a~\citet{Kroupa2001} initial mass function (IMF), the solar metallicity
CCSN yields of~\citet{Chieffi2013} and~\citet[][based on
the~\citealp{Sukhbold2016} models with forced explosion]{Griffith2021a}
predict~$\ycc{O} = 0.016$ and 0.018, respectively, if all stars
from~$8 - 120~\msun$ explode.
The value of~$\ycc{O} = 0.015$ allows for a modest amount of black hole
formation but implicitly assumes that most massive stars explode.\footnote{
	If all stars from~$8 - 40~\msun$ explode and all more massive stars
	collapse, then the~\citet{Sukhbold2016} models with forced explosions
	yield~$\ycc{O} = 0.013$~\citep{Griffith2021a}.
}
In chemical evolution models, this choice of yield also leads to good agreement
with the observationally inferred deuterium-to-hydrogen ratio of the local
ISM~\citep{Linsky2006}, while substantially lower~$\ycc{O}$ leads to
disagreement~\citep{Weinberg2017b}.
\par
Our adopted values of~\ycc{O}~and~\ycc{Fe}~give~$\ofe\approx0.43$ for stars
with pure CCSN enrichment, in good agreement with the ``high-$\alpha$'' plateau
of disk stars found by~\citet{Ramirez2013}; matching the APOGEE plateau
at~$\ofe\approx0.35$~\citep[see, e.g., fig.~6 of][]{Hasselquist2021} would
instead require a slightly higher~$\ycc{Fe} = 0.0014$.
SN Ia models predict minimal O yields, justifying~$\yia{O} = 0$.
The choice of~$\yia{Fe} = 0.00214$ then leads to good agreement with the
observed~\ofe~values of low-$\alpha$ thin-disc stars given the star formation
assumptions used by~\citet{Johnson2021} (for analytic discussion, see~\S~3.1
of~\citealp*{Weinberg2017}).
For an Fe yield of~$0.77~\msun$ from a single SN Ia event~\citep{Iwamoto1999},
this~\yia{Fe}~corresponds to a time-integrated SN Ia rate
of~$R_\text{Ia} = 2.7\times10^{-3}~\msun^{-1}$ (i.e., 2.7 SNe Ia per
1000~\msun~of star formation), which is moderately higher than the value
of~$2.2\times10^{-3}~\msun^{-1}$ inferred by~\citet{Maoz2012} for
a~\citet{Kroupa2001} IMF.
Our choice of yields is internally consistent and reproduces many Milky Way
observations~\citep{Johnson2021}, but many of the GCE model predictions would
be minimally affected if we lowered~\ycc{O},~\ycc{Fe}, and~\yia{Fe}~by a
common factor and reduced the efficiency of outflows.
We return to this point in the context of N yields
in~\S~\ref{sec:results:yields:variations}.
\par
We assume that N is not produced in significant amounts by SNe Ia
\citep{Johnson2019}, setting~$\yia{N} = 0$.
The remainder of this section discusses the CCSN and AGB star yields of N.
\par
A significant portion of N yields arise as a consequence of the CNO
cycle.\footnote{
	\Ctwelve(p,~$\gamma$)\Nthirteen($\beta^+~\nu_\text{e}$)\Cthirteen
	(p,~$\gamma$)\Nfourteen(p,~$\gamma$)\Ofifteen($\beta^+~\nu_\text{e}$)
	\Nfifteen(p,~$\alpha$)\Ctwelve
}
As the dominant source of pressure and energy generation in non-zero
metallicity main sequence stars with initial masses of~$\gtrsim$1.3~\msun, this
cyclic nuclear reaction catalyses the conversion of H into He that would
otherwise be accomplished by the proton-proton chain
(\citealp{vonWeizsaecker1937, vonWeizsaecker1938, Bethe1939a, Bethe1939b,
Adelberger2011};~\citealp*{Suliga2021}).
Its slowest component by far is the~\Nfourteen(p,~$\gamma$)\Ofifteen~reaction
\citep[e.g.][]{LUNA2006}.
Consequently, the first order effect of the CNO cycle is to convert most of the
C isotopes in stellar cores into~\Nfourteen.
As we will discuss in this section, this plays an important role in shaping N
yields from stars of all masses.

\subsection{Core Collapse Supernovae and Massive Star Winds}
\label{sec:yields:ccsne}

In~\vice, CCSN nucleosynthetic products are approximated to be produced
instantaneously following an episode of star formation; this is a good
approximation because the lives of massive stars are short compared to the
relevant timescales for GCE.
The yield is simply the constant of proportionality between the CCSN production
rate and the star formation rate (SFR):
\begin{equation}
\dot{M}_\text{X}^\text{CC} = \ycc{X}\dot{M}_\star
\end{equation}
More generally,~\ycc{X}~quantifies~\textit{all} of the nucleosynthetic material
approximated to be produced instantaneously following a single stellar
population's formation, including newly synthesized material expelled in a
massive star wind before the star explodes or collapses to a black hole.
\par
We compute theoretically predicted values of~\ycc{N}~using
\vice's~\texttt{vice.yields.ccsne.fractional} function assuming a
\citet{Kroupa2001} IMF; details on how~\vice~handles these calculations can be
found in~\S~4 of~\citet{Griffith2021a} and in the~\vice~science 
documentation\footnote{\url
	{https://vice-astro.readthedocs.io/en/latest/science_documentation/yields}
}.
In the left panel of Fig.~\ref{fig:n_cc_yields}, we plot the results as a
function of progenitor metallicity as predicted by the~\citet{Woosley1995},
\citet*{Nomoto2013},~\citet{Sukhbold2016}, and~\citet{Limongi2018} tables.
There is generally good agreement between the various non-rotating models, but
only~\citet{Limongi2018} report yields for progenitors with non-zero rotational
velocities; these yields are substantially larger than their non-rotating
counterparts, especially at low metallicity.
With few seed nuclei for the CNO cycle at low~$Z$, production of~\Nfourteen~is
difficult.
Rotation-induced mixing, a highly uncertain process~\citep{Zahn1992, Maeder1998,
Lagarde2012}, could transport newly produced C and O into the hydrogen burning
shell of the CCSN progenitor, facilitating~\Nfourteen~production
(\citealp{Frischknecht2016}; see also discussion in~\S~4.2 of
\citealp{Andrews2017}).
Consequently, N yields at low metallicity are quite sensitive to model-dependent
assumptions regarding stellar rotation and internal mixing processes
\citep{Heger2010}.
\par
We compute the~\no~ratio of CCSN ejecta from the values
of~\ycc{N}~and~\ycc{O}~predicted by a given yield table according to
\begin{equation}
\no\subcc = 
\log_{10}\left(\frac{\ycc{N}}{\ycc{O}}\right) -
\log_{10}\left(\frac{Z_{\text{N},\odot}}{Z_{\text{O},\odot}}\right),
\label{eq:no_subcc}
\end{equation}
where~$Z_{\text{X},\odot}$ is the abundance by mass of some element X in the
sun, for which we take~$Z_{\text{N},\odot} = 6.91\times10^{-4}$ and
$Z_{\text{O},\odot} = 5.72\times10^{-3}$ based on the photospheric measurements
of~\citet{Asplund2009}.
For each value of~\ycc{N}~in the left panel of Fig.~\ref{fig:n_cc_yields}, we
compute the corresponding values of~\ycc{O}~and illustrate the
resultant~\no\subcc~ratios in the right panel.
These yield ratios follow similar trends with progenitor metallicity and
rotation as~\ycc{N}~itself, a consequence of the fact that these
studies predict relatively metallicity- and rotation-independent O yields.
At low metallicity, CCSN yields of N dominate over the AGB star yields (see
discussion in~\S~\ref{sec:yields:agb}), and Fig.~\ref{fig:no_oh_observed}
suggests a plateau in~\no~at low metallicity at~$\no\subcc\approx-0.7$.
Taking this value in combination with our adopted O yield of~$\ycc{O} = 0.015$,
equation~\refp{eq:no_subcc} suggests that~$\ycc{N} = 3.6\times10^{-4}$.
We highlight both~\no\subcc~=~$-0.7$ and~$\ycc{N} = 3.6\times10^{-4}$ with
horizontal black dashed lines in Fig.~\ref{fig:n_cc_yields}, finding good
agreement with the rotating progenitor models of~\citet{Limongi2018} in both
panels.
This indicates that rotating massive stars play an important role in
establishing the N abundances at low metallicity, in agreement with previous
works~\citep{Chiappini2003, Chiappini2005, Chiappini2006, Kobayashi2011,
Prantzos2018, Grisoni2021}.
We therefore take~$\ycc{N} = 3.6\times10^{-4}$ as our fiducial CCSN yield of N;
both the normalization and metallicity-independence of this choice are
supported by the~\citet{Limongi2018} models.

\begin{figure*}
\centering
\includegraphics[scale = 0.45]{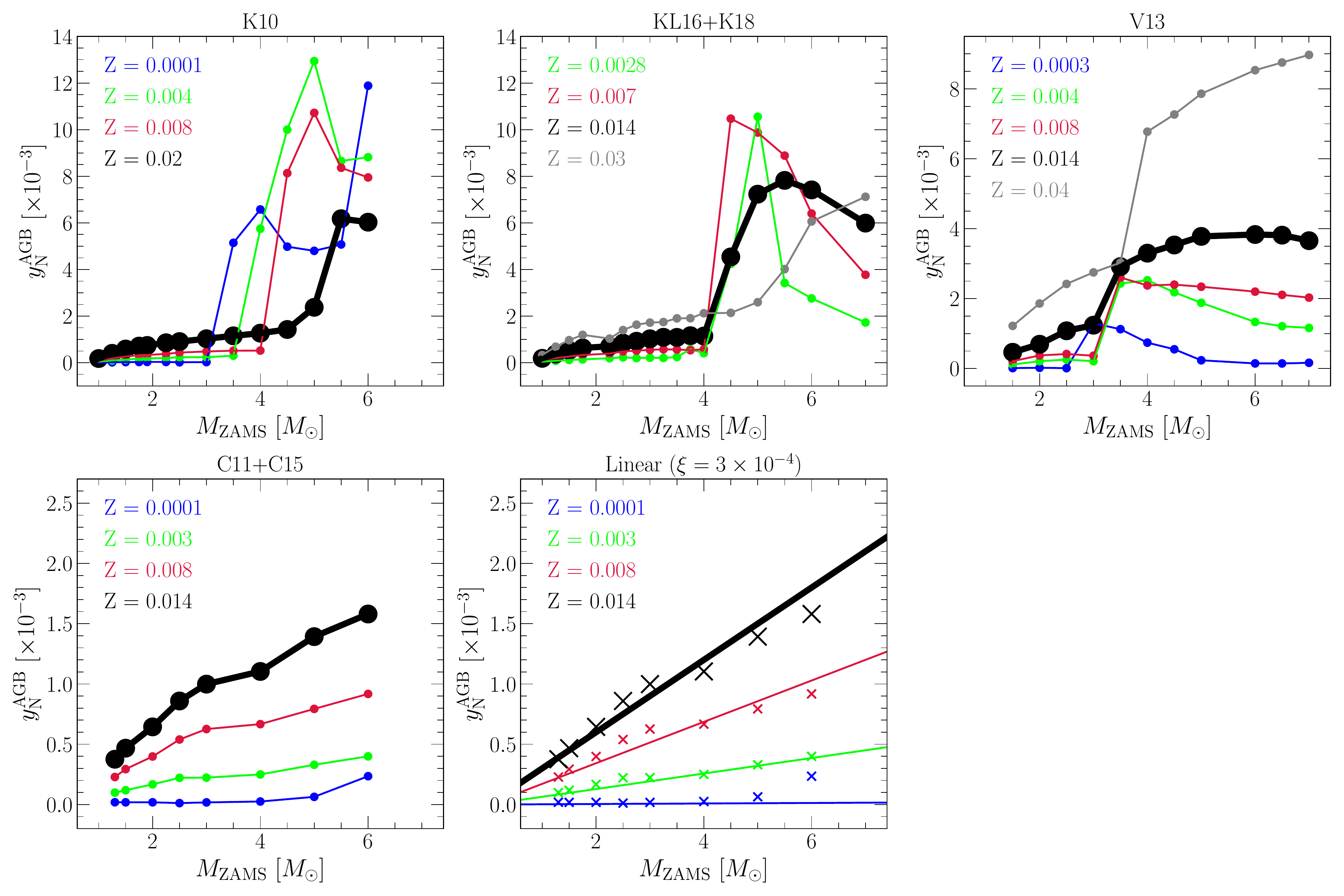}
\caption{
The fractional yields of N from AGB stars~\yagb{N}~as a function of progenitor
ZAMS mass and birth metallicity~$Z$ as reported by
\citet[][upper left]{Karakas2010},~\citet{Karakas2016} and
\citet[][upper middle]{Karakas2018},~\citet[][upper right]{Ventura2013,
Ventura2014, Ventura2018, Ventura2020}, and~\citet[][lower left]{Cristallo2011,
Cristallo2015}.
For~\citet{Ventura2013, Ventura2014, Ventura2018, Ventura2020} and
\citet{Cristallo2011, Cristallo2015}, we show the yields only for a selection
of metallicities available from their provided tables.
We highlight yields at solar metallicity ($Z = 0.02$ for~\citealp{Karakas2010},
$Z = 0.014$ otherwise) with bold black lines.
In the lower right panel, we show our linear model (coloured lines, see
equation~\ref{eq:linear_yield}) in comparison to
the~\citet[][coloured X's]{Cristallo2011, Cristallo2015} predictions.
We caution that the y-axis ranges are not the same between panels in this
figure.
}
\label{fig:agb_yield_models}
\end{figure*}

The~\citet{Sukhbold2016} tables, available only at solar metallicity, agree
nearly perfectly with our empirical value of~$\ycc{N} = 3.6\times10^{-4}$, but
they predict a higher value of~\no\subcc~by~$\sim$0.2 dex.
This is a consequence of the failed supernovae incorporated into their model
and the lowered values of~\ycc{O} that result (see discussion
in~\S~\ref{sec:results:yields}).
While N emerges in substantial amounts in winds, much of the O produced by
massive stars is ejected during the explosion, making the O yield more
sensitive to the black hole landscape~\citep{Griffith2021a}.
Most of the SN models plotted in Fig.~\ref{fig:n_cc_yields} estimate slightly
higher~\no\subcc~at~$\log_{10}(Z / Z_\odot)$ = 0 relative to our empirical
value of~\no\subcc~=~$-0.7$, but they still fall short of solar~\no.
This implies the need for an additional enrichment channel, which is expected
because it is well understood that N is also produced in considerable amounts
by AGB stars~\citep[e.g.][]{Johnson2019}.

\subsection{Asymptotic Giant Branch Stars}
\label{sec:yields:agb}

Similar to SNe, our AGB star yields are parametrized as fractional net yields.
For a yield~$\yagb{X}$, the mass yield is given by~$M_\star \yagb{X}$.
Enrichment proceeds as it does in~\citet{Johnson2021}: AGB stars place their
nucleosynthetic products in the~$\delta\rgal = 100$ pc ring that they are in at
a given time, allowing stars to enrich distributions of radii as they migrate.
\vice~implements an algorithm that computes the mass in dying stars
from each stellar population, and the zero age main sequence (ZAMS) mass
required to compute the fractional yield comes from a mass-lifetime
relationship. 
For the latter, we adopt the metallicity-independent parabola in
$\log\tau - \log m$ space from~\citet{Larson1974} with updated coefficients
from~\citet{Kobayashi2004} and~\citet*{David1990} (see discussion of the
mass-lifetime relationship in~\vice~in Appendix~\ref{sec:vice}).
\par
We make use of four previously published tables of AGB star N yields
computed from stellar evolution models, each of which are sampled on a grid
of progenitor masses and metallicities.
To approximate the net yield~\yagb{X}~as a smooth function of~$M_\star$ and
$Z_\star$,~\vice~interpolates bi-linearly -- once in mass~$M$ and once in
metallicity~$Z$ -- and linearly extrapolates above or below the grid in either
quantity as necessary.
By comparing the predicted abundances of the~\citet{Johnson2021} Milky Way
model to the latest observational data, we can constrain how accurately these
``off-the-shelf'' yield models characterize N production.
These models taken from the literature are as follows.
\begin{itemize}
	\item[\textbf{1.}] \citet[][hereafter~\karakasten]{Karakas2010}\footnote{
		We clarify that our abbreviations of each of these papers refer
		specifically to their yields of N as we adopt them in our model.
		We cite the full names of these papers when referring to their more
		general results.
	} published yields for~$Z = 0.0001$, 0.004, 0.008, and 0.02 progenitors.
	We plot these yields in the upper left panel of 
	Fig.~\ref{fig:agb_yield_models}.

	\item[\textbf{2.}] \citet{Karakas2016} and~\citet{Karakas2018} published
	yields for~$Z = 0.0028$, 0.007, 0.014, and 0.03 progenitors; we hereafter
	refer to these yields as the~\karakas~model.
	We illustrate these yields in the upper middle panel of
	Fig.~\ref{fig:agb_yield_models}.

	\item[\textbf{3.}] We combine the yields for~$Z = 0.0003$ and 0.008
	progenitors from~\citet{Ventura2013} with those at~$Z = 0.004$ from
	\citet{Ventura2014}, at~$Z = 0.014$ from~\citet{Ventura2018}, and at
	$Z = 0.04$ from~\citet{Ventura2020} into a single table of yields.
	In this set, we also include a set of un-published yields at~$Z = 0.001$
	and 0.002 computed from similar models (provided by P. Ventura, private
	communication).
	We hereafter refer to this yield set as the~\ventura~model, and we plot a
	subsample of these yields in the upper right panel of
	Fig.~\ref{fig:agb_yield_models}.

	\item[\textbf{4.}] The default set of AGB star yields in~\vice~is taken
	from~\citet{Cristallo2011, Cristallo2015}, who published yields for
	$Z = 0.0001$, 0.0003, 0.001, 0.002, 0.003, 0.006, 0.008, 0.01, 0.014, and
	0.02 progenitors.
	We hereafter refer to these yields as the~\cristallo~model, and we
	illustrate a subsample of them in the lower left panel of
	Fig.~\ref{fig:agb_yield_models}.
\end{itemize}
\par
\vice~also allows users to construct their own functions of progenitor mass
and metallicity to describe the AGB star yield.
Motivated by the roughly linear nature of the~\cristallo~yields and their
general success once renormalized by a constant factor (see discussion
in~\S~\ref{sec:results:yields}), we construct a model in which the yield is
linearly proportional to both progenitor ZAMS mass and metallicity:
\begin{equation}
\yagb{N} = \xi \left(\frac{M_\star}{M_\odot}\right)
\left(\frac{Z_\star}{Z_\odot}\right).
\label{eq:linear_yield}
\end{equation}
We illustrate this model in the lower middle panel of Fig.
\ref{fig:agb_yield_models} for~$\xi = 3\times10^{-4}$ in comparison to
the~\cristallo~yields shown by the coloured X's.
Although we find good agreement between the~\cristallo~yields and our linear
model with a normalization of~$\xi = 3\times10^{-4}$, for our fiducial AGB star
yield of N we take a slope of~$\xi = 9\times10^{-4}$.
We discuss the absolute scaling of our nucleosynthetic yields
in~\S~\ref{sec:results:yields} below.
\par
As is clear from Fig.~\ref{fig:agb_yield_models}, the N yields reported by
these studies show substantial differences.
Unfortunately, ascertaining the origin of these differences is difficult
because each model employs its own assumptions for important evolutionary
parameters such as opacity, mass loss, nuclear reaction networks, and
convection and convective boundaries within stars, all of which have a
significant impact on stellar evolution and thus the predicted yields (see
discussion in, e.g.,~\S~5 of~\citealp{Karakas2016}).
However, the differences can be qualitatively understood by considering two
important phenomena known to occur within AGB stars: TDU\footnote{
	The time adverbial ``third'' in TDU refers only to the fact that these
	dredge-up episodes are occurring while the star is on the asymptotic giant
	branch. Because they are associated with the thermal pulsations of AGB
	stars, there are many episodes of third dredge-up.
} and HBB.
The variations in how TDU and HBB proceed between different stellar evolution
models arise as consequences of the different input physics.
\par
When an AGB star experiences a thermal pulsation, this is usually accompanied
by a TDU event whereby the convective envelope penetrates into the
hydrogen-depleted core, mixing some of this material with other material
exposed to partial He-shell burning.
The~\Cthirteen($\alpha$, n)\Osixteen~reaction, which is the main source of free
neutrons in low-mass AGB stars~\citep{Gallino1998}, can occur at substantial
rates when this core material is mixed with the He-rich shell.
This process does not directly affect N abundances in the shell because the
core is mostly composed of C and O at this evolutionary phase,
but~\Nfourteen~plays an important role in shaping an AGB star's overall
\textit{s}-process yield by acting as an efficient catalyst of neutron decay
via the~\Nfourteen(n, p)\Cfourteen($\beta^+~\nu_{e}$)\Nfourteen~reaction, the
first step of which is a resonant neutron capture~\citep{Cristallo2011}.
\par
HBB refers to proton capture reactions at the base of the convective envelope,
activating the CNO cycle and producing large amounts of~\Nfourteen~at the
expense of C and O isotopes.
HBB requires a higher mass AGB star progenitor ($M_\text{ZAMS} = 4 - 5~M_\odot$
at~$Z_\odot$ according to~\citealt{Karakas2010}) than TDU
($M_\text{ZAMS} = 2 - 2.5~M_\odot$ at~$Z_\odot$ according to
\citealt{Karakas2010}), but the minimum mass for both decreases at lower
metallicities.
\par
The most efficient N production occurs when both TDU and HBB are active within
an AGB star, because each replenishment of C and O isotopes by TDU adds new
seed nuclei for the CNO cycle with HBB.
This is the reason for the substantial increase in yields at~$\sim$4~\msun~in
the~\karakasten~and~\karakas~models; in both yield sets, every star that
experiences HBB also experiences TDU (see, e.g., table 1
of~\citealp{Karakas2010}).
Their high mass AGB star yields are higher at low~$Z$ because both HBB and TDU
are more efficient~\citep[see discussion in][]{Ventura2013}: when the
metallicity is low, each TDU episode is deeper due to the lower opacity, and
the base of the convective envelope is hotter, increasing the rate of CNO cycle
reactions in HBB.
This interaction between TDU and HBB is also the reason for the increase in
the~\ventura~yields near~$\sim$3~\msun, but unlike
the~\karakasten~and~\karakas~models, their stars experience both processes only
in this narrow range of mass.
\par
Of all of these yields taken from the literature, the~\cristallo~sample shows
the smoothest dependence on progenitor mass and metallicity.
Below~$\sim$3~\msun, their agreement with the~\karakas~yields is good, but this
model has much lower N yields for higher mass AGB stars.
Pinpointing a single reason for this difference is difficult, even when
considering the differences between HBB and TDU.
Relative to the~\karakas~yields (see discussion in~\S~5 of
\citealp{Karakas2016}), the~\cristallo~stars have more mass loss, fewer
thermal pulses overall, and weaker HBB due to a lower temperature at the base
of the convective envelope, each of which acts to lower the yield of~\Nfourteen.
\par
Although the~\karakasten~and~\karakas~yield models both show a substantial
increase in N yields above~$\sim$4~\msun, there are some noteworthy differences
between the two.
In the newer version, the yields at solar metallicity are somewhat higher,
and the yields at sub-solar metallicities decreased slightly, particularly for
the highest mass AGB stars.
These differences can be understood by slight variations in the input physics
(A. Karakas, private communication).
A portion of the increase in the yields at solar metallicity can be attributed
to the assumption of~$Z_\odot = 0.02$ versus~$Z_\odot = 0.014$\footnote{
	Changes in the accepted value of the metallicity of the sun trace back to
	the canonical value of~$\sim$2\% derived by, e.g.,~\citet{Anders1989} and
	\citet{Grevesse1998}, later being revised to~$\sim$1.4\% by, e.g.,
	\citet{Lodders2003} and~\citet*{Asplund2005}. See Table 4 of
	\citet{Asplund2009} for a compilation of measured values.
} and the impact this has on both HBB and TDU, but it does not account for the
entire difference.
As a consequence of updates to opacity tables and the adopted solar composition,
the~\karakas~models at solar metallicity are slightly hotter and more compact,
giving them hotter HBB and deeper TDU.
With more thermal pulses overall and therefore a longer AGB lifetime, these
stars have more time to convert~\Ctwelve~into~\Nfourteen.
\karakas~also use low-temperature opacity tables based on~\citet{Marigo2002}
that more closely follow the surface composition of the star.
These opacities are high, making the more massive AGB stars larger and
increasing the mass-loss rate relative to~\karakasten, truncating their N
yields.
The~$Z = 0.0028$ model uses the~\citet{Bloecker1995} mass-loss prescription
rather than that of~\citet{Vassiliadis1993}, which was used for
the~\karakasten~yields as well as the yields at other metallicities in
the~\karakas~model.
This choice results in fewer thermal pulses and a shorter AGB lifetime, giving
them less time to process C and O nuclei into~\Nfourteen.
\par
In the interest of consistency, when we adopt a particular AGB star yield model
for N, we also adopt the corresponding table within~\vice~for O and Fe when
possible.\footnote{
	In the case of~\citet{Ventura2013, Ventura2014, Ventura2018, Ventura2020},
	AGB star yields of Fe are not available, and our linear model is only
	appropriate for N.
	In these cases, we assume the~\vice~default of the~\citet{Cristallo2011,
	Cristallo2015} yields for both O and Fe.
}
These O and Fe yields, however, are negligible compared to their SN yields.
Although we focus our investigation on AGB yields from~$\lesssim$~7~\msun~stars,
slightly more massive stars (up to~$\sim$12~\msun) sit near the critical mass
boundaries between different types of massive white dwarfs and electron capture
SN progenitors.
\citet{Doherty2017} investigated theoretically predicted yields of these stars
and found significant production of CNO isotopes.
There is also the intriguing possibility of the CNO yields from the earliest,
most metal-poor AGB stars (e.g. the~$Z = 10^{-5}$ models
of~\citealp{Gil-Pons2013, Gil-Pons2021}) and the insight this may afford into
N production at low~$Z$ and the most metal-poor stars in the Galaxy.
While experiments with such yields in our GCE models would be interesting, this
is beyond the scope of the current paper since our AGB yield models already
span a wide range of assumptions regarding stellar evolution.

\subsection{IMF-Averaged AGB Star Yields: Metallicity and Time Dependence}
\label{sec:yields:imf_agb}

\begin{figure*}
\centering
\includegraphics[scale = 0.46]{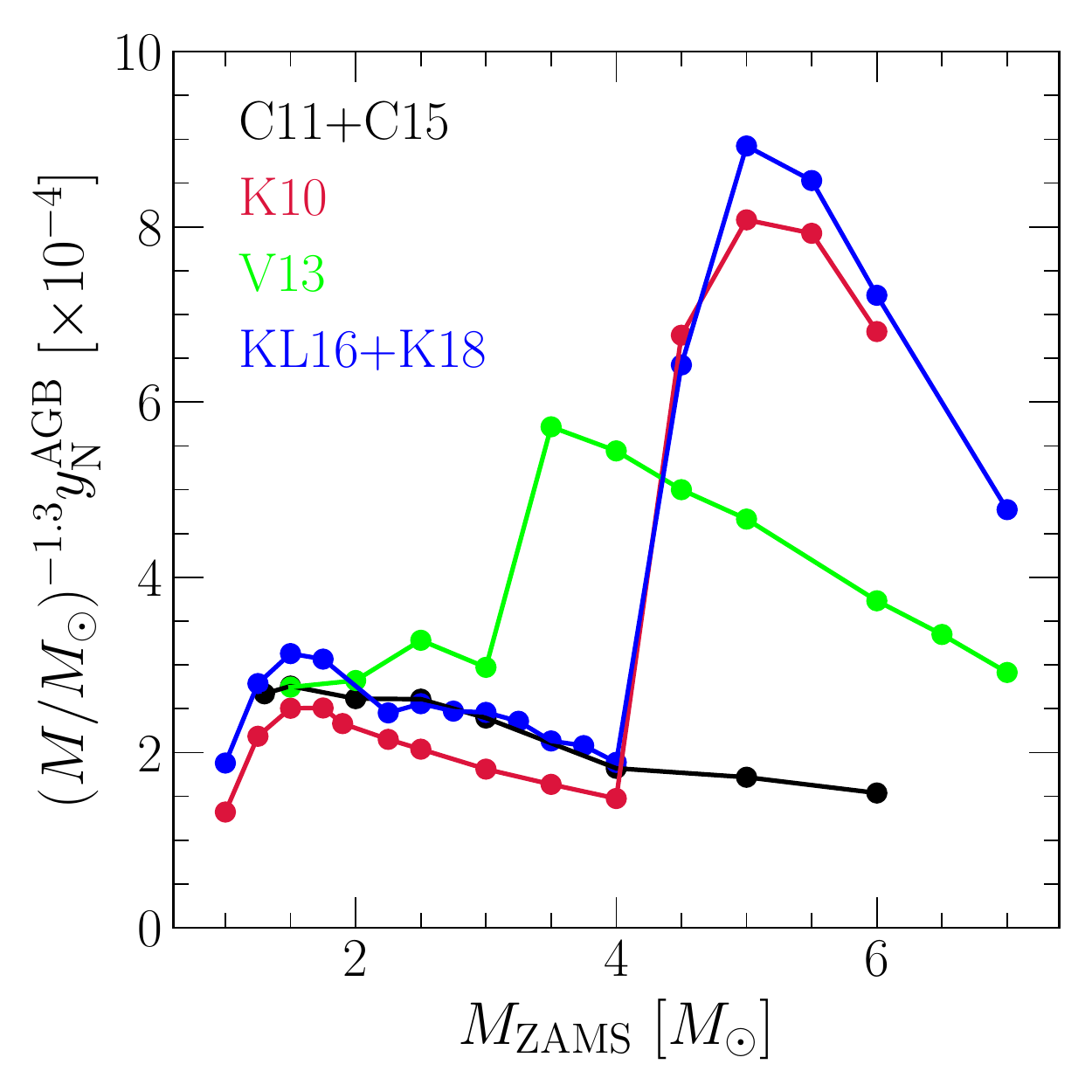}
\includegraphics[scale = 0.45]{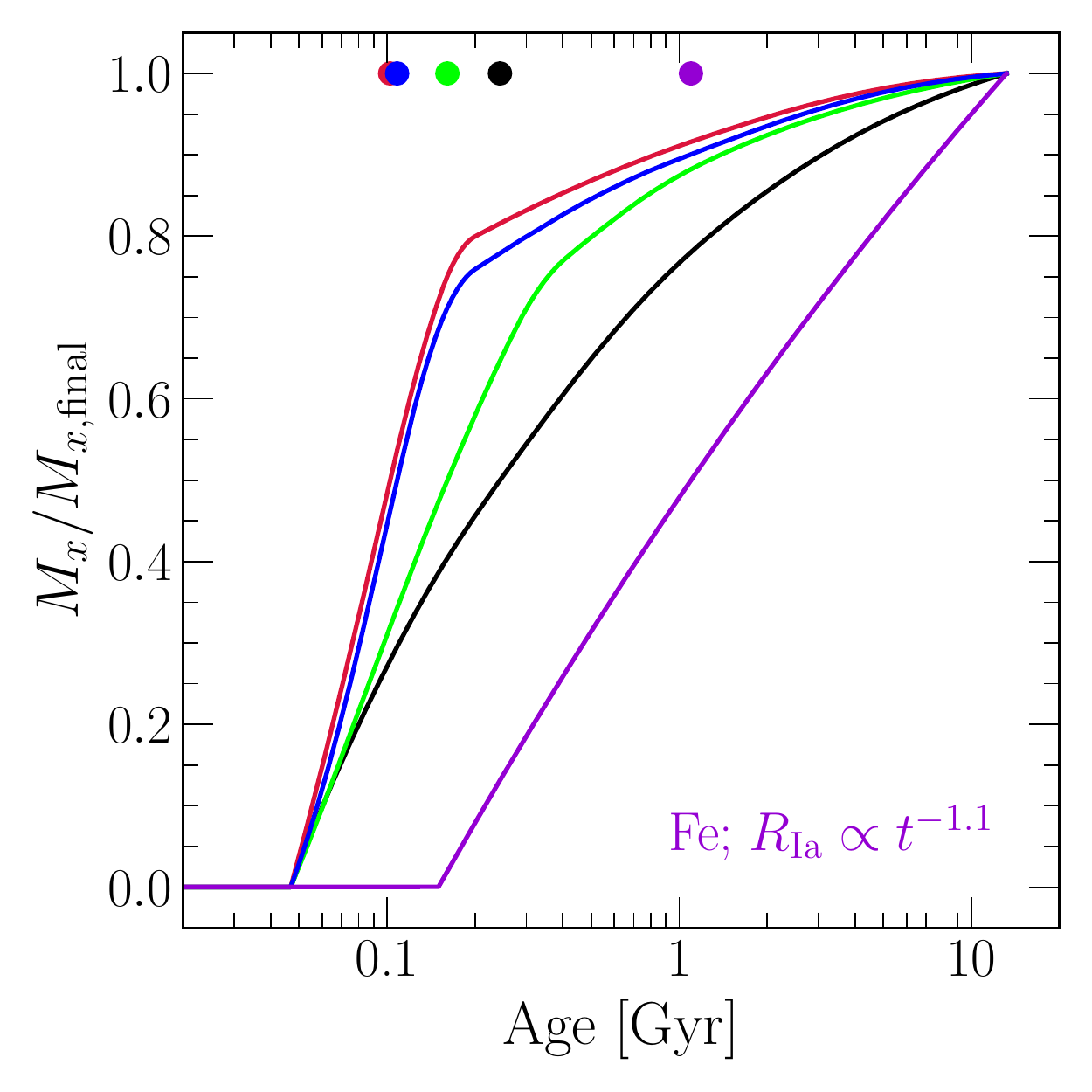}
\includegraphics[scale = 0.46]{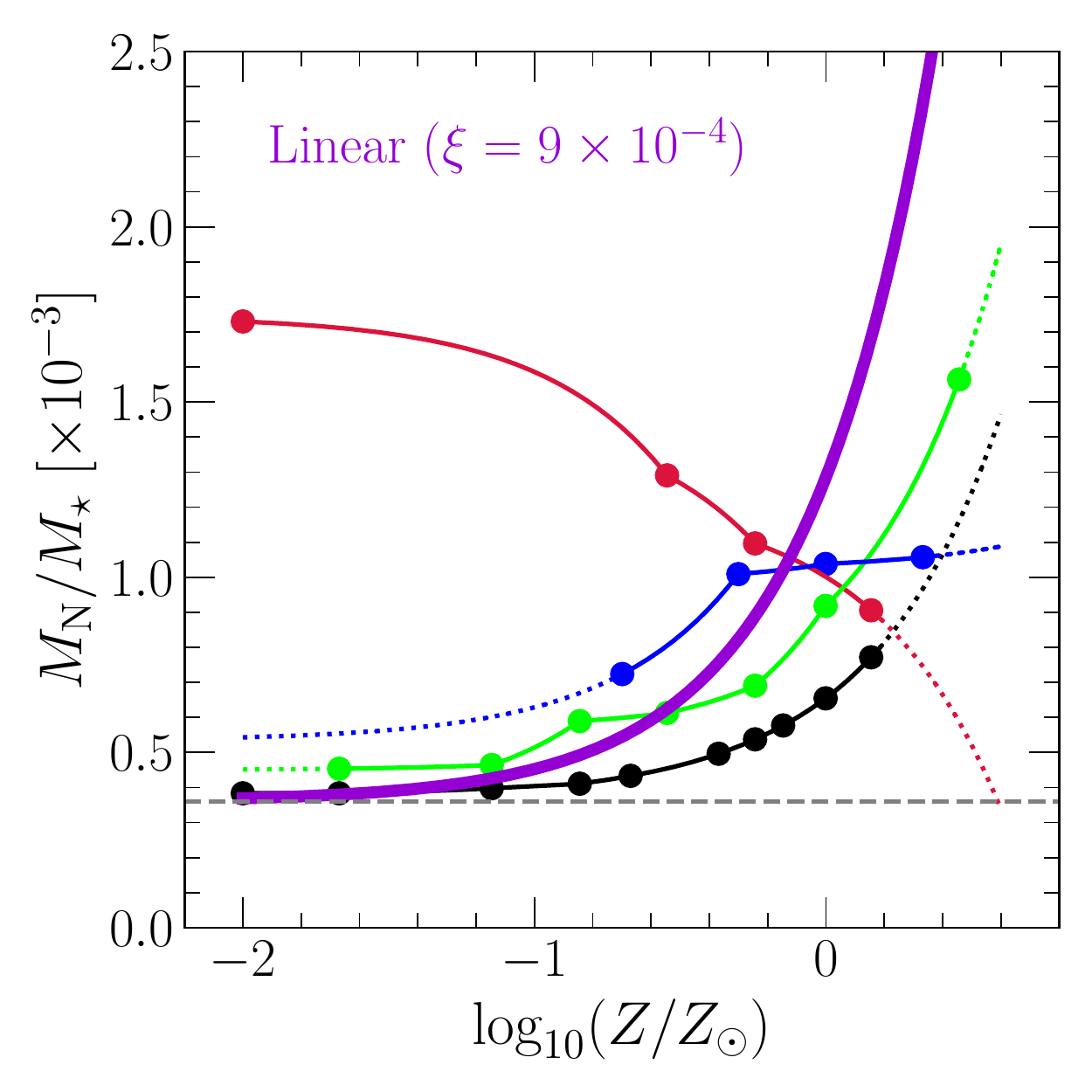}
\caption{
\textbf{Left}: The IMF-weighted mass yield of N from AGB stars as a function of
progenitor ZAMS mass at solar metallicity (i.e. the contribution per linear
interval~dM$_\text{ZAMS}$;~$Z_\odot = 0.014$).
\textbf{Middle}: The net mass of N produced by AGB stars from a single stellar
population for each of our yield models at solar metallicity.
The purple line denotes the same for Fe assuming our~$t^{-1.1}$ DTD as in the
\citet{Johnson2021} chemical evolution model.
All values are normalized to the total mass produced at an age of 13.2 Gyr.
Points at the top of the panel denote the ages at which 50\% of the total mass
yield has been produced.
\textbf{Right}: The total amount of N produced by a 13.2 Gyr old stellar
population as a function of metallicity for each of our yield models normalized
by the stellar population's initial mass.
Points mark metallicities at which the published tables report yields, and the
lines are dotted at metallicities that are above (below) the maximum (minimum)
metallicity reported by a given study (i.e. where extrapolation is necessary).
In this panel only, we include the metallicity-independent contribution
$\ycc{N} = 3.6\times10^{-4}$ from CCSNe (gray dashed line).
The bold purple curve represents our inference of the~\textit{total} N yield
(CCSN + AGB) required to reproduce the observational constrains discussed
in~\S~\ref{sec:results} given our adopted O and Fe yields (\S~\ref{sec:yields}).
}
\label{fig:ssp}
\end{figure*}

To more directly compare these AGB star yields predicted from stellar
evolution models, we plot their IMF-weighted yields at solar metallicity in the
left hand panel of Fig.~\ref{fig:ssp}.
We assume~$Z_\odot = 0.014$ based on~\citet{Asplund2009}
and~\citet*{Asplund2021}; since the~\karakasten~model reports yields at
$Z = 0.02$ rather than~$Z = 0.014$, we simply interpolate linearly
to~$Z = 0.014$ in the same manner that~\vice~does in our GCE models.
As mentioned in~\S~\ref{sec:yields:agb}, the AGB star yield~\yagb{N}~as we
have parametrized it is in units of the progenitor star's ZAMS mass, and
consequently the~\textit{mass yield} of N is given by~$M_\star \yagb{N}$.
With an additional weight of~$M_\star^{-2.3}$ from the IMF in this mass range
\citep[e.g.][]{Kroupa2001}, we therefore multiply the values of~\yagb{N}~by
$(M_\star / M_\odot)^{-1.3}$ to quantify a star's relative contribution to the
total N yield taking into account the intrinsic mass distribution.\footnote{
	This weight gives a contribution per linear interval of M$_\text{ZAMS}$, so
	one can use area under the curve to assess relative contributions.
}
With the additional weight of~$M_\star^{-1.3}$, the~\cristallo~yields are
relatively mass-independent.
For the other yield models, higher mass AGB stars dominate the overall yield
due to the effects of TDU and HBB discussed in~\S~\ref{sec:yields:agb}.
\par
Using~\vice's~\texttt{vice.single\_stellar\_population} function, in the
middle panel of Fig.~\ref{fig:ssp} we plot the total N yield as a function of
age from a single stellar population.
For the sake of this calculation, we set all CCSN yields of N to zero in order
to highlight the AGB star contribution.
We show the results of this procedure for solar metallicity only, and we
normalize all values
to the total mass produced at~$t = 13.2$ Gyr (the total amount of time our GCE
model is integrated over; see discussion in~\S~\ref{sec:multizone}).
Under the~\cristallo~yields, it takes~$\sim$250 Myr for a single stellar
population to produce~$\sim$50\% of its N from AGB stars, as noted by the
coloured points at the top of the panel.
This is in good agreement with~\citet{Maiolino2019}, who find that similar
parameter choices predict 80\% of the N yield to be ejected within~$\sim$1 Gyr
(see their fig. 1).
The characteristic timescales for N production are even shorter in the other
yield models because of their more pronounced contributions from massive stars
with short lifetimes~\citep[e.g.][]{Larson1974, Maeder1989, Padovani1993}.
For comparison, we plot the enrichment of Fe by our~$t^{-1.1}$ power-law DTD,
also with the CCSN yield set to zero to highlight the delayed component.
The characteristic delay time for Fe production is considerably longer than
that of N -- up to an order of magnitude depending on which yield model is
adopted.
As noted in~\citet{Johnson2021}, a characteristic delay time of~$\sim$1 Gyr
is exactly as expected for a~$\sim t^{-1}$ DTD because half of the SN Ia
events occur between 100 Myr and 1 Gyr and the other half between 1 Gyr and
10 Gyr.


A characteristic delay time of only~$\sim$250 Myr may seem surprising given
the relatively mass-independent nature of the IMF-weighted~\cristallo~yields.
This arises out of the steep nature of the stellar mass-lifetime relation
\citep[e.g.][]{Larson1974, Maeder1989, Padovani1993}.
For example, 2 and 3~\msun~stars live only~$\sim$1.2 Gyr and~$\sim$400 Myr,
respectively, and over the course of 13.2 Gyr, only stars of masses
$\gtrsim$0.9~\msun~will have enough time to finish their hydrogen burning.
Consequently, most of the mass range of stars that will evolve through an
AGB phase will do so within the first few hundred Myr after their formation,
and with mass-independent IMF-weighted yields, this accounts for most of the
N.
We clarify that the delay times computed here apply~\textit{only} to N and
not necessarily to other elements produced by AGB stars.
As we have illustrated here, the effective DTD of AGB star enrichment is
dictated by the combination of the stellar mass-lifetime relation and the
mass dependence of the yield, which should in principle differ from element to
element.
Other elements produced by slow neutron capture often have the highest yields
from lower mass AGB stars.
For example,~\citet{Cristallo2011, Cristallo2015} report Sr yields that are
dominated by~$M_\text{ZAMS} = 2 - 3~\msun$ progenitors (see fig. 5 of
\citealp{Johnson2020}), giving it a characteristic delay time of~$\sim$500 Myr.
The characteristic delay-times will be as long as a few Gyr if and only if the
yields are dominated by~$\lesssim$1.5~\msun~stars.
\par
In the right panel of Fig.~\ref{fig:ssp}, we plot the total amount of N
produced by a 13.2 Gyr old single stellar population as a function of its
initial metallicity according to all of our AGB star yield tables, including
the linear model (see equation~\ref{eq:linear_yield} and discussion
in~\S~\ref{sec:yields:agb}).
For this calculation, we include the metallicity-independent CCSN yield
($\ycc{N} = 3.6\times10^{-4}$; see discussion in~\S~\ref{sec:yields:ccsne}).
In general, there is good qualitative agreement between the~\cristallo~and
the~\ventura~models, the only major difference being the normalization.
The predictions with the linear model with~$\xi = 3\times10^{-4}$ are nearly
identical to the~\cristallo~model, as one would expect from
Fig.~\ref{fig:agb_yield_models}, but here we show the yields for our fiducial
choice of~$\xi = 9\times10^{-4}$.
The value at which these N yields flatten off at low~$Z$ is reflective of our
adopted value of~\ycc{N} (grey dashed line).
Up to~$\log_{10}(Z / Z_\odot) \approx -0.2$, the~\karakas~yields predict a
similar trend as~\cristallo~and~\ventura, also with a difference in
normalization, but at solar and super-solar metallicities they predict much
more metallicity-independent N yields than others.
The~\karakasten~yields, on the other hand, do not agree with any of the other
models, instead predicting N yields to~\textit{decrease} monotonically with
increasing~$Z$.
These differences between the~\karakasten~and~\karakas~models trace back to
differences regarding the opacity and mass loss prescriptions (see discussion
in~\S~\ref{sec:yields:agb}).
Although the normalization depends on the SN yields of all elements, we
demonstrate in~\S~\ref{sec:results:yields} that reproducing the~\ohno~relation
as observed requires AGB N yields which scale roughly linearly with metallicity
as in the~\cristallo~and~\ventura~models.
More specifically, with our adopted O and Fe yields (see discussion at the
beginning of~\S~\ref{sec:yields}), reproducing the observational constraints
that we consider requires~\textit{total} N yields (CCSN + AGB) with the
metallicity dependence shown by the purple curve in the right panel of
Fig.~\ref{fig:ssp}.

\section{The Multi-Zone Chemical Evolution Model}
\label{sec:multizone}

We use the fiducial model for the Milky Way published by
\citet{Johnson2021}, which runs using the~\vice~GCE code (see Appendix
\ref{sec:vice};~\citealp{Johnson2020, Griffith2021a}).
Multi-zone models allow simultaneous calculations of abundances for multiple
Galactic regions, making them a more physically realistic option than classical
one-zone models for a system like the Milky Way.
Furthermore, they can take into account stellar migration in a framework that
is much less computationally expensive than hydrodynamical simulations, making
them the ideal experiments for our purposes.
We provide a brief summary of the model here, but a full breakdown can be found
in~\S~2 of~\citet{Johnson2021}.
\par
As in previous models for the Milky Way with similar motivations
\citep[e.g.][]{Matteucci1989, Schoenrich2009, Minchev2013, Minchev2014,
Minchev2017, Sharma2021}, this model parametrizes the Galaxy disc as a series
of concentric rings.
With a uniform width of~$\delta\rgal = 100$ pc, each ring is assigned its own
star formation history (SFH), and with assumptions about outflows and the
$\Sigma_\text{gas}-\dot{\Sigma}_\star$ relation (see discussion
below),~\vice~computes the implied amounts of gas and infall at each timestep
automatically.
Under the caveat that stellar populations can move and place some of their
nucleosynthetic products in rings other than the one they were born in, each
ring is otherwise described by a conventional one-zone GCE model.
Allowing stars to enrich distributions of radii was a novel addition to this
type of model, and~\citet{Johnson2021} demonstrated that this has a
significant impact on enrichment rates from delayed sources such as SNe Ia.
\par
To drive stellar migration, the model makes use of star particles from a
hydrodynamical simulation, for which~\citet{Johnson2021} select the~\hsim~galaxy
from the~\citet{Christensen2012} suite evolved with the N-body+SPH code
\texttt{GASOLINE}~\citep{Wadsley2004}; we retain this decision here.
\hsim~spans 13.7 Gyr of evolution, but the sample of star particles with
reliable birth radii span 13.2 Gyr in age; the model thus places the onset of
star formation~$\sim$500 Myr after the Big Bang and integrates up to the
present day.
Previous studies have shown that~\hsim, among other disc galaxies evolved with
similar physics, has a realistic rotation curve~\citep{Governato2012,
Christensen2014a, Christensen2014b}, stellar mass~\citep{Munshi2013},
metallicity~\citep{Christensen2016}, dwarf satellite population
\citep{Zolotov2012, Brooks2014}, HI properties~\citep{Brooks2017}, and stellar
age-velocity relation~\citep{Bird2021}.
Despite this, there are some interesting differences between~\hsim~and the
Milky Way.
First and foremost,~\hsim~had only a weak and transient bar and lacks one at
the present day, while the Milky Way is known to have a strong, long-lived
central bar~\citep[e.g.][]{Bovy2019}.
This could indicate that the dynamical history of~\hsim~and its star particles
differs significantly from that of the Milky Way.
Furthermore, the last major merger in~\hsim~was at a redshift of~$z \approx 3$,
making it an interesting case study for its quiescent merger history
\citep[e.g.][]{Zolotov2012}, while the Sagitarrius dwarf galaxy is believed to
have made pericentric passages around the Milky Way at~$1 - 2$ Gyr
intervals~\citep{Law2010}.
Although these differences between~\hsim~and the Milky Way are well understood,
their impact on chemical evolution is not.
We are unaware of any studies that investigate the impact of different
assumptions regarding the Galaxy's dynamical history on predicted abundances;
this is however an interesting question for future work.
\par
Radial migration of stars proceeds from the~\hsim~star particles in a simple
manner; for a stellar population in our model born at a radius~$R_\text{birth}$
and a time~$t_\text{birth}$,~\vice~searches for star particles born at
$R_\text{birth} \pm 250$ pc and $t_\text{birth} \pm 250$ Myr.
From the star particles that make this cut, it then randomly selects one to act
as that stellar population's~\textit{analogue}.
The stellar population then assumes the present day midplane distance~$z$ and
the change in orbital radius~$\Delta\rgal$ of its analogue between its birth
and the present day.
In the~\citet{Johnson2021} fiducial model, stellar populations move to their
implied final radii with a~$\sqrt{\text{age}}$ dependence according to:
\begin{equation}
\rgal(T) = R_\text{birth} + \Delta\rgal
\sqrt{\frac{t - t_\text{birth}}{13.2~\text{Gyr} - t_\text{birth}}},
\end{equation}
where 13.2 Gyr is simply the present day (see discussion above).
With displacement proportional to~$\sqrt{\text{age}}$, this corresponds to a
scenario in which radial migration proceeds as a diffusion process as modeled
by~\citet{Frankel2018, Frankel2019} and supported by the N-body simulations
of~\citet*{Brunetti2011}.
Although~\citet{Johnson2021} investigated other assumptions for this
time-dependence, in the present paper we only use this parametrization
(hereafter referred to as the ``diffusion'' prescription) and an idealized one
in which stars remain at their birth radius until they instantaneously migrate
at the present day (hereafter referred to as the ``post-processing''
prescription).
If~\vice~does not find any star particles from~\hsim~in its initial
$\rgal \pm 250$ pc and~$t \pm 250$ Myr search, it widens it to
$\rgal \pm 500$ pc and~$t \pm 500$ Myr; if still no candidate analogues are
found,~\vice~maintains the~$t \pm 500$ Myr requirement, but assigns the star
particle with the smallest difference in birth radius as the analogue.
This procedure can be thought of as ``injecting'' the dynamics of
the~\hsim~galaxy into our multi-zone chemical evolution model, and it can in
principle be repeated for any hydrodynamical simulation of a disc galaxy.
As in~\citet{Johnson2021}, we neglect radial gas flows
\citep[e.g.][]{Lacey1985, Bilitewski2012, Vincenzo2020}, instead focusing on
the impact of stellar migration.
\par
Rather than using a hydrodynamical simulation, some previous studies have
implemented stellar migration using dynamical arguments
\citep[e.g.][]{Schoenrich2009, Sharma2021}.
An advantage of our approach over this is that these dynamical arguments
introduce free parameters which then require fitting to data.
It is also unclear to what extent the fit might bias the model into agreement
with quantities in the sample not involved in the fit.
A disadvantage of our approach is that we are restricted to one realization of
our dynamical history; slight variations are not possible.
We do not distinguish between ``blurring'' and ``churning'', terms commonly
used to refer to changes in the guiding centre radii of stars and their
epicyclic motions, respectively.
Both are induced by a variety of physical interactions such as molecular cloud
scattering~\citep{Mihalas1981, Jenkins1990, Jenkins1992}, orbital resonances
with spiral arms or bars~\citep{Sellwood2002, Minchev2011}, and satellite
perturbations~\citep*{Bird2012}.
All of these effects are included in~\hsim~and should therefore be inherited by
the stellar populations in our GCE model.
\par
We assume the SFH of the ``inside-out'' model from~\citet{Johnson2021}.
The time-dependence at a given~\rgal~is described by
\begin{equation}
\dot{\Sigma}_\star \propto (1 - e^{-t / \tau_\text{rise}})
e^{-t/\tau_\text{sfh}},
\end{equation}
where~$\tau_\text{rise}$ approximately controls the amount of time the SFR is
rising at early times; we set this parameter equal to 2 Gyr at all radii as in
\citet{Johnson2021}.
Our e-folding timescales of~$\tau_\text{sfh}$ are taken from a fit of this
functional form to the~$\Sigma_\star$-age relation in bins of~$R / R_\text{e}$
for~$10^{10.5} - 10^{11}~M_\odot$ Sa/Sb Hubble type spiral galaxies reported
by~\citet{Sanchez2020}.
The resulting values of~$\tau_\text{sfh}$ are long:~$\sim$15 Gyr at the solar
circle (\rgal~= 8 kpc) and as high as~$\sim$40 Gyr in the outer disc (see fig.
3 of~\citealp{Johnson2021}).
This is a consequence of flat nature of the~$\Sigma_\star$-age relation
reported by~\citet{Sanchez2020}.
\par
Within each~$\delta\rgal = 100$ pc ring, the normalization of the SFH is set by
the total stellar mass of the Milky Way disc and the present-day stellar
surface density gradient, assuming that it is unaffected by migration (see
Appendix B of~\citealp{Johnson2021}).
For the former, we neglect the contribution from the bulge and adopt the total
disc stellar mass of~$5.17\times10^{10}~\msun$ from~\citet{Licquia2015}.
For the latter, we adopt a double exponential form describing the thin- and
thick-disc components.
We take the scale radii of the thin- and thick-discs to be~$R_\text{t} = 2.5$
kpc and~$R_\text{T} = 2.0$ kpc, respectively, with a surface density ratio
at~\rgal~= 0 of~$\Sigma_\text{T} / \Sigma_\text{t} = 0.27$ based on the
findings of~\citet{Bland-Hawthorn2016}.

Since the~\citet{Johnson2021} models run~\vice~with the SFH specified a priori,
determining the gas supply requires an assumption regarding the SFE.
Based on the findings of~\citet{Kennicutt1998}, GCE models often adopt a
single power-law relating the surface density of gas~$\Sigma_\text{gas}$ to
the surface density of star formation~$\dot{\Sigma}_\star$.
Recent studies, however, have revealed that the spatially resolved
$\Sigma_\text{gas} - \dot{\Sigma}_\star$ relation is more nuanced than the
integrated relation~\citep{delosReyes2019, Ellison2021, Kennicutt2021}.
Some of the uncertainty regarding its details can be traced back to the
ongoing debate about the CO-to-H$_2$ conversion factor
(\citealp{Kennicutt2012};~\citealp*{Liu2015}).
Based on a compilation of the~\citet{Bigiel2010} and~\citet{Leroy2013} data
shown in comparison to the theoretically motivated parametrizations of
\citet[][see their fig. 2]{Krumholz2018},~\citet{Johnson2021} take a
three-component power-law~$\dot{\Sigma}_\star \propto \Sigma_\text{gas}^N$ with
index~$N$ given by:
\begin{equation}
N =
\begin{cases}
1.0 & (\Sigma_\text{gas} \geq 2\times10^7~\msun~\persqkpc) \\
3.6 & (5\times10^6~\msun~\persqkpc \leq \Sigma_\text{gas} \leq
2\times10^7~\msun~\persqkpc) \\
1.7 & (\Sigma_\text{gas} \leq 5\times10^6~\msun~\persqkpc).
\end{cases}
\label{eq:sflaw_indeces}
\end{equation}
The normalization is fixed by setting the SFE timescale
$\tau_\star \equiv \Sigma_\text{gas} / \dot{\Sigma}_\star$ at surface densities
where~$N = 1$ to the value derived observationally for molecular gas, denoted
by~$\tau_\text{mol}$.
This value at the present day is taken to be~$\tau_{\text{mol},0} = 2$ Gyr
\citep{Leroy2008, Leroy2013}, with a~$t^{1/2}$ dependence on cosmic time based
on the findings of~\citet{Tacconi2018} studying the properties of molecular gas
as a function of redshift.
\par
With the yields adopted in the~\citet{Johnson2021} models (see discussion at
the beginning of~\S~\ref{sec:yields}), considerable outflows are required in
order to predict empirically plausible abundances.
\citet{Weinberg2017} demonstrate analytically that the equilibrium abundance of
some element in the gas phase is approximately determined by its yield and the
mass loading factor~$\eta \equiv \dot{\Sigma}_\text{out} / \dot{\Sigma}_\star$
with a small correction for the SFH.
\citet{Johnson2021} select a scaling of~$\eta$ with~\rgal~such that the
equilibrium abundance as a function of radius corresponds to a reasonable
metallicity gradient within the Galaxy (see their fig. 3 and discussion in
their~\S~3.1).
Nonetheless, yields and the strength of outflows are mutually degenerate
parameters since they act as source and sink terms in computing enrichment
rates.
The absolute scale of nucleosynthetic yields is a topic of debate (see
discussion in, e.g.,~\citealp{Griffith2021a}), and some authors even neglect
outflows entirely, arguing that they do not significantly alter the chemical
evolution of the Galaxy disc~\citep[e.g.][]{Spitoni2019, Spitoni2021}.
We investigate the impact of simultaneous variations in our yields and the
efficiency of outflows in~\S~\ref{sec:results:yields} below.

\section{Results}
\label{sec:results}

In this section, we present the predictions of our GCE models.
We establish a fiducial model in~\S~\ref{sec:results:fiducial}, which adopts
$\ycc{N} = 3.6\times10^{-4}$ and our linear AGB star yields (equation
\ref{eq:linear_yield}) with~$\xi = 9\times10^{-4}$ along with the O and Fe
yields of~\citet[see discussion in~\S~\ref{sec:yields}]{Johnson2021}.
We discuss the evolution of this model in radius and time as well as the impact
of stellar migration.
In~\S~\ref{sec:results:yields}, we consider AGB star yield models taken from
the literature (see discussion in~\S~\ref{sec:yields:agb}) and use an
empirical~\ohno~relation to discriminate among them.
For each of these previously published yields, we explore alternate
parametrizations of our SN yields and/or the efficiency of outflows which may
address their shortcomings in reproducing the observed trend.
We return to the fiducial model in~\S~\ref{sec:results:t_z_dep_comp} to
demonstrate the dominance of the metallicity dependence of N yields over the
AGB star DTD in establishing the gas-phase~\ohno~relation.
To make the comparisons more clear, we make use of the post-processing
migration prescription in~\S\S~\ref{sec:results:yields}
and~\ref{sec:results:t_z_dep_comp} but otherwise retain the diffusion
prescription (see discussion in~\S~\ref{sec:multizone}).
In~\S~\ref{sec:results:vincenzo_comp}, we compare our fiducial~\no~vs. age
and~\no~vs.~\ofe~trends to the stellar abundances derived from APOGEE data
by~\citet{Vincenzo2021}.
In~\S~\ref{sec:results:schaefer_comp}, we demonstrate how variations in the
SFE or inflow rate can induce scatter in the gas-phase~\ohno~relation.
We finish presenting our results in~\S~\ref{sec:results:ohno_equilibrium} by
offering an analytic understanding of the~\ohno~relation obtained with
chemical equilibrium arguments inspired by~\citet{Weinberg2017}.
As a reference for the reader, in Table~\ref{tab:resultsref} we provide a
summary of which migration prescriptions we take as a default and which AGB
star yield models we consider in each subsequent subsection.

\subsection{Evolution of a Fiducial Model}
\label{sec:results:fiducial}

\begin{figure*}
\centering
\includegraphics[scale = 0.6]{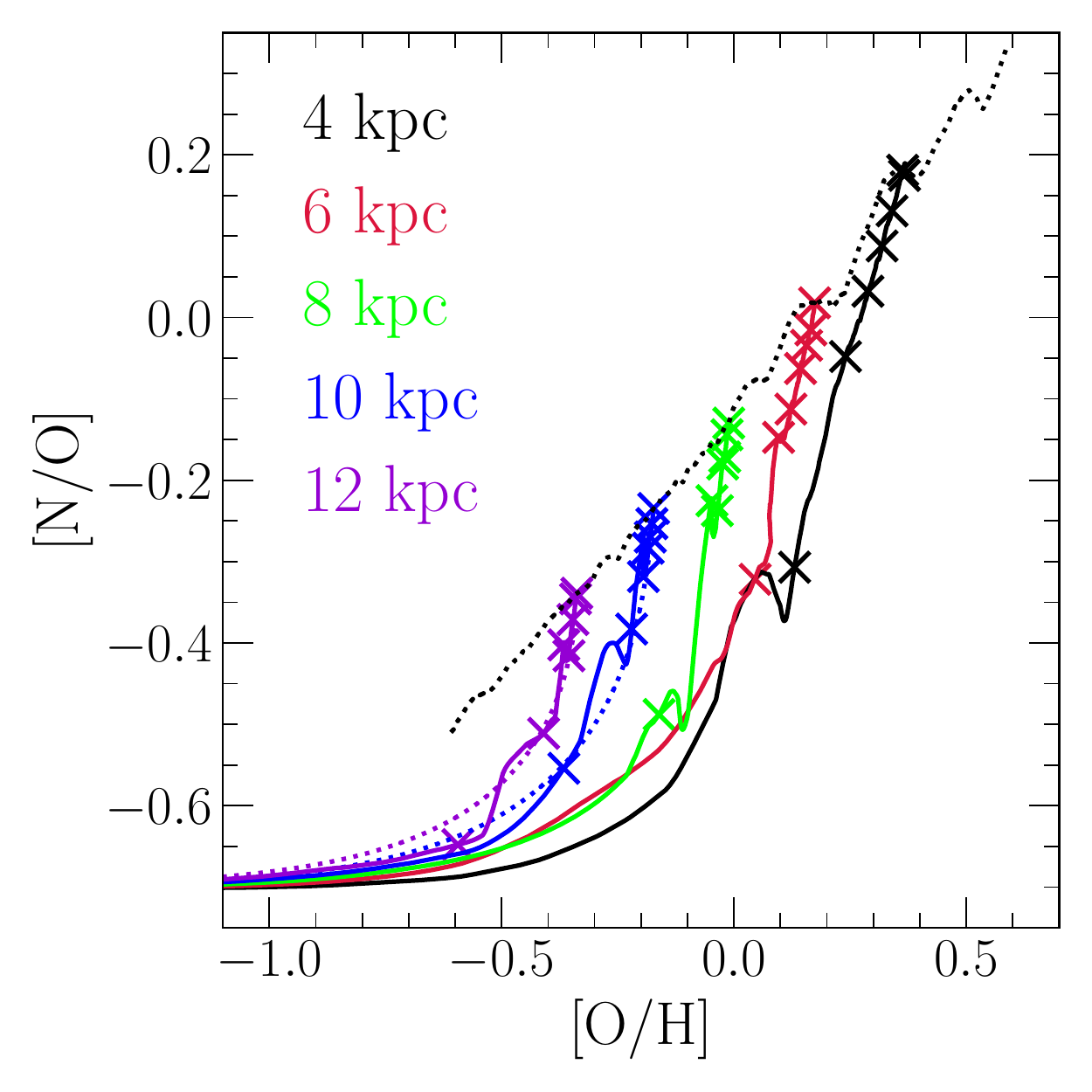}
\includegraphics[scale = 0.6]{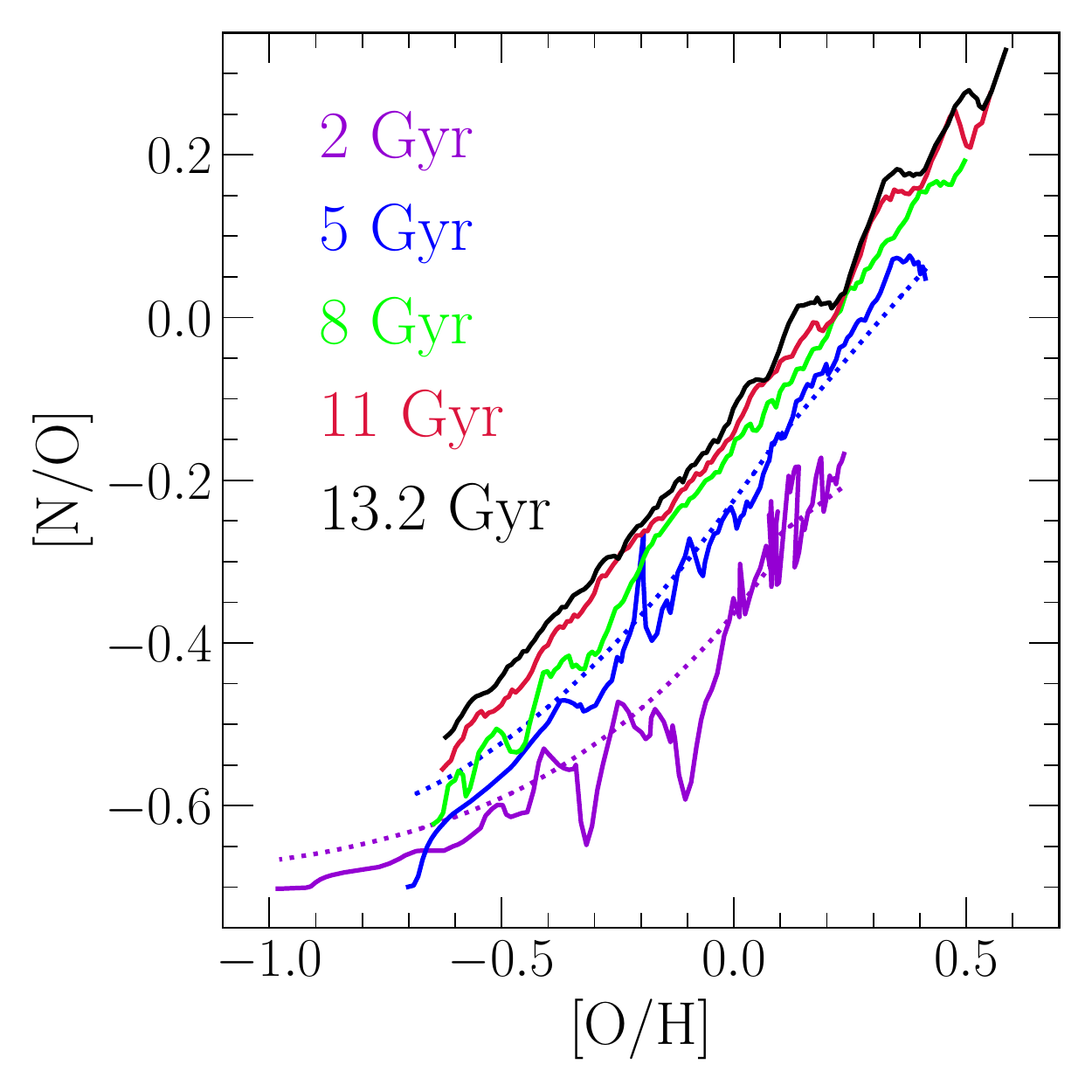}
\caption{
\textbf{Left}: The gas-phase~\ohno~relation parametrized by time at
fixed radius (solid coloured lines) in the fiducial model. X's denote the
abundances at~$t = 2$, 4, 6, 8, 10, 12, and 13.2 Gyr (the present day) at all
radii.
The dotted black line is the same as the solid black line in the right panel.
Coloured dotted lines mark the evolution of our model at~$\rgal = 10$ and 12
kpc when we neglect the impact of stellar migration on enrichment rates (i.e.
the ``post-processing'' migration prescription from~\citealp{Johnson2021}; see
discussion in~\S~\ref{sec:multizone}).
\textbf{Right}: The gas-phase~\ohno~relation parametrized by radius at
various snapshots (solid coloured lines) in our fiducial model.
Similar to the left panel, coloured dotted lines denote the resulting relation
at~$t = 2$ and 5 Gyr when we neglect stellar migration in computing enrichment
rates.
}
\label{fig:no_oh_timeevol}
\end{figure*}

\par
In the left panel of Fig.~\ref{fig:no_oh_timeevol}, we plot the evolution of N
and O abundances in the gas phase at five different Galactocentric radii
in our fiducial model (see paragraph above).
At early times,~\oh~is low and~\no~reflects the ratio of the CCSN yields
(\no\subcc~$\approx -0.7$).
Consequently, the tracks in each ring are similar.
Once lower mass stars begin to evolve through an AGB phase, they enrich the
ISM with N but negligible amounts of O, increasing~\no.
At this point, the tracks in each ring separate from one another.
This separation is a consequence of the metallicity gradient in~\oh~being
established early in the Galaxy's evolution.
The radial gradient in our model arises out of a decrease in the equilibrium
abundance of O with increasing radius.
Produced on short timescales by CCSNe, O achieves equilibrium faster than
elements produced by delayed nucleosynthetic sources~\citep{Weinberg2017}.
The ISM therefore reaches equilibrium in O soon after AGB stars begin
producing N, after which~\no~continues to increase at an approximately
fixed~\oh~at all radii (see also Fig.~\ref{fig:nh_feh_vs_lookback} and
associated discussion in~\S~\ref{sec:results:vincenzo_comp}).
The separation of these evolutionary tracks contests the popular interpretation
that the~\ohno~relation arises as an evolutionary sequence, instead suggesting
a superposition of evolutionary endpoints at equilibrium values set by
different outflow efficiencies.
Similar arguments have been made regarding low [$\alpha$/Fe] disc
stars~\citep[e.g.][]{Schoenrich2009, Nidever2014, Buck2020, Sharma2021}.
In short, a given ring in our model does not evolve along the~\ohno~relation,
instead following this ``rightward-then-upward'' trajectory dictated by the
timescales on which N and O achieve equilibrium.

\begin{table}
\caption{
A reference on which migration prescription (see~\S~\ref{sec:multizone}) we
take as a default and which AGB star yield model(s) we consider in each
subsection of~\S~\ref{sec:results}.
Our linear AGB star yield is defined in equation~\refp{eq:linear_yield}.
We use a metallicity-independent IMF-averaged massive star yield of
$\ycc{N} = 3.6\times10^{-4}$ throughout, exploring alternate,
metallicity-dependent parametrizations in~\S~\ref{sec:results:yields:yncc} only
(right panel of Fig.~\ref{fig:no_oh_predictions}).
We do not present multi-zone GCE models
in~\S\S~\ref{sec:results:yields:summary} and~\ref{sec:results:ohno_equilibrium}.
}
\begin{tabularx}{\columnwidth}{l @{\extracolsep{\fill}} l r}
\hline
Section & AGB Star Yield Model(s) & Migration Prescription
\\
\hline
\ref{sec:results:fiducial} & Linear &
Diffusion
\\
\ref{sec:results:yields} & \cristallo,~\ventura,~\karakasten,~\karakas &
Post-processing
\\
\ref{sec:results:yields:variations} & Linear,~\cristallo,~\ventura &
Post-processing
\\
\ref{sec:results:yields:yncc} & \karakasten,~\karakas & Post-processing
\\
\ref{sec:results:yields:summary} & N/A & N/A
\\
\ref{sec:results:t_z_dep_comp} & Linear & Post-processing
\\
\ref{sec:results:vincenzo_comp} & Linear & Diffusion
\\
\ref{sec:results:schaefer_comp} & Linear & Diffusion
\\
\ref{sec:results:ohno_equilibrium} & N/A & N/A
\\
\hline
\end{tabularx}
\label{tab:resultsref}
\end{table}

Because there is a delay between a stellar population's formation and N
production from its AGB stars ($\sim$250 Myr in this model; see Fig.
\ref{fig:ssp}), stellar migration can in principle occur within this time
interval.
Although the bulk of migration occurs on longer timescales, this characteristic
delay is comparable to the dynamical time of the Milky Way and is thus adequate
for kinematic heating to at least begin.
In zoom-in hydrodynamical simulations from the FIRE\footnote{
	\url{https://fire.northwestern.edu}
} suite~\citep{Hopkins2014},~\citet{El-Badry2016} find that stars in a
M$_\star \approx 10^{10.6}~\msun$ galaxy can migrate~$1 - 2$ kpc within 1 Gyr
of their formation.
Consequently, N enrichment rates at fixed~\rgal~may differ significantly from
their expected values given the SFH at that radius because stellar migration
induced a deficit or surplus of N-producing AGB stars.
These tracks can thus move vertically in the~\ohno~plane in response to AGB
stars moving between rings as the Galaxy evolves, producing the ``jitter'' in
these evolutionary tracks.
We demonstrate this effect by comparing the solid blue and purple lines to
their dotted counterparts.
These are the tracks we compute using the post-processing migration
prescription which eliminates the impact of migration on enrichment rates (see
discussion in~\S~\ref{sec:multizone}).
\par
In the right panel of Fig.~\ref{fig:no_oh_timeevol}, we plot the
gas-phase~\ohno~relation predicted by the model at various snapshots.
To obtain this, we simply take the N and O abundances in the ISM at a given
output time for each~$\delta\rgal = 100$ pc ring at~$\rgal > 2$ kpc and plot
them as a line.
The relation is generally time-independent at~$t \gtrsim 5$ Gyr.
Although there is some slight evolution toward higher~\no, the total change
in~\no~over this time interval is well within the intrinsic scatter derived
observationally (see Fig.~\ref{fig:no_oh_observed}).
Even at~$t = 2$ Gyr, corresponding to~$z \approx 2.6$,~\no~at fixed~\oh~is
only~$\sim$0.2 dex lower than its value at the present day.
Especially when considering the intrinsic scatter that would arise if we were
to consider models with, e.g., different SFHs, this calculation supports
previous arguments that the redshift evolution of the~\ohno~relation is
minimal~\citep{Vincenzo2018, HaydenPawson2021}.
\par
We again demonstrate the impact of stellar migration in the right panel of
Fig.~\ref{fig:no_oh_timeevol} by comparing the blue and purple solid lines
to their dotted counterparts, which quantify the relation using the
post-processing migration prescription.
This indicates that the local jitter seen in the relation at a given time is
a consequence of migration as discussed above.
The mechanism by which stellar migration imposes these features in
the~\ohno~plane is qualitatively similar to what~\citet{Johnson2021} find for
SN Ia production of Fe.
They found that the SN Ia rate in this model can vary by as much as a factor of
$\sim$3 at large radii ($\rgal \gtrsim 9$ kpc).
When a deficit or surplus of SN Ia events is sustained for timescales
comparable to the depletion time of the local ISM, the gas-phase abundance of
Fe increases or decreases accordingly.
As a consequence, some of the stellar populations that form during these
events are Fe-poor enough to present as young stars ($\lesssim$ 6 Gyr) with
significantly super-solar [$\alpha$/Fe] ratios, some of which are indeed
observed in the solar neighbourhood with APOGEE\footnote{
	Although some of these stars appear to have mis-labeled ages
	\citep{Jofre2016, Yong2016, Izzard2018, SilvaAguirre2018, Miglio2021}, this
	phenomenon could explain an intrinsically young sub-component
	(\citealp{Hekker2019}; see discussion in~\S\S~3.1 and 3.4 of
	\citealp{Johnson2021}).
}~\citep{Chiappini2015, Martig2015, Martig2016, Warfield2021}.
In the case of N, the effect is smaller ($\lesssim$0.1 dex in~\no) because our
model predicts N yields to be ejected from stellar populations~$\sim$5
times faster than Fe (Fig.~\ref{fig:ssp}).
Consequently, there is less time for stellar migration to occur within the
timescale of N production than there is within the timescale of Fe production.
\par
To assess how well our model characterizes N production in the Milky Way and
external galaxies, in the next section we compare the
present-day~\ohno~relation (solid black line in the right panel of
Fig.~\ref{fig:no_oh_timeevol}) to the~\citet{Dopita2016} trend.
We do the same for our AGB yield models taken from the literature
(see discussion in~\S~\ref{sec:yields:agb}), exploring variations of other
model parameters which may address their shortcomings.
We then return to our fiducial model thereafter for additional tests against
observed data.

\subsection{Comparison to Observed Gas-Phase Trends}
\label{sec:results:yields}

\begin{figure*}
\centering
\includegraphics[scale = 0.45]{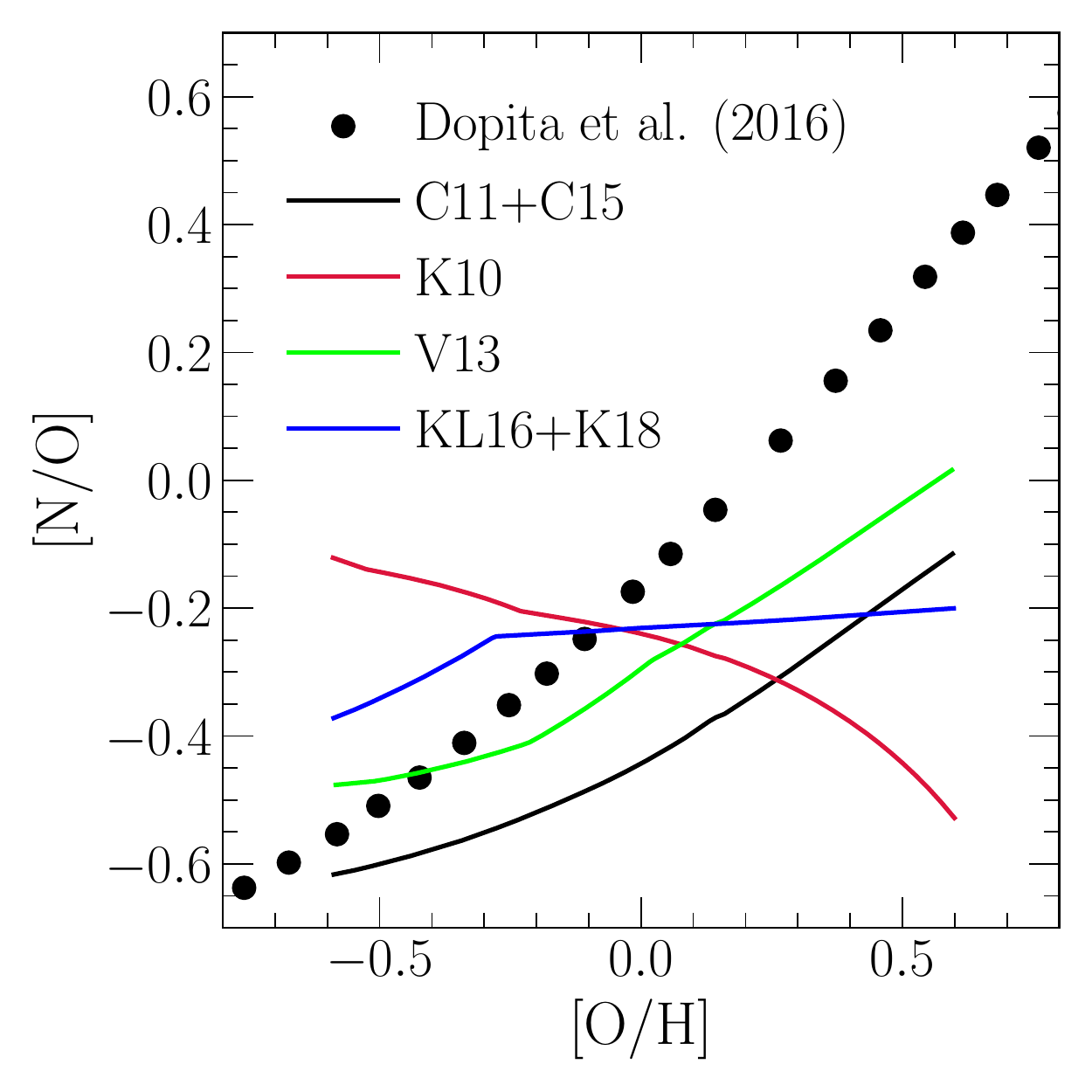}
\includegraphics[scale = 0.45]{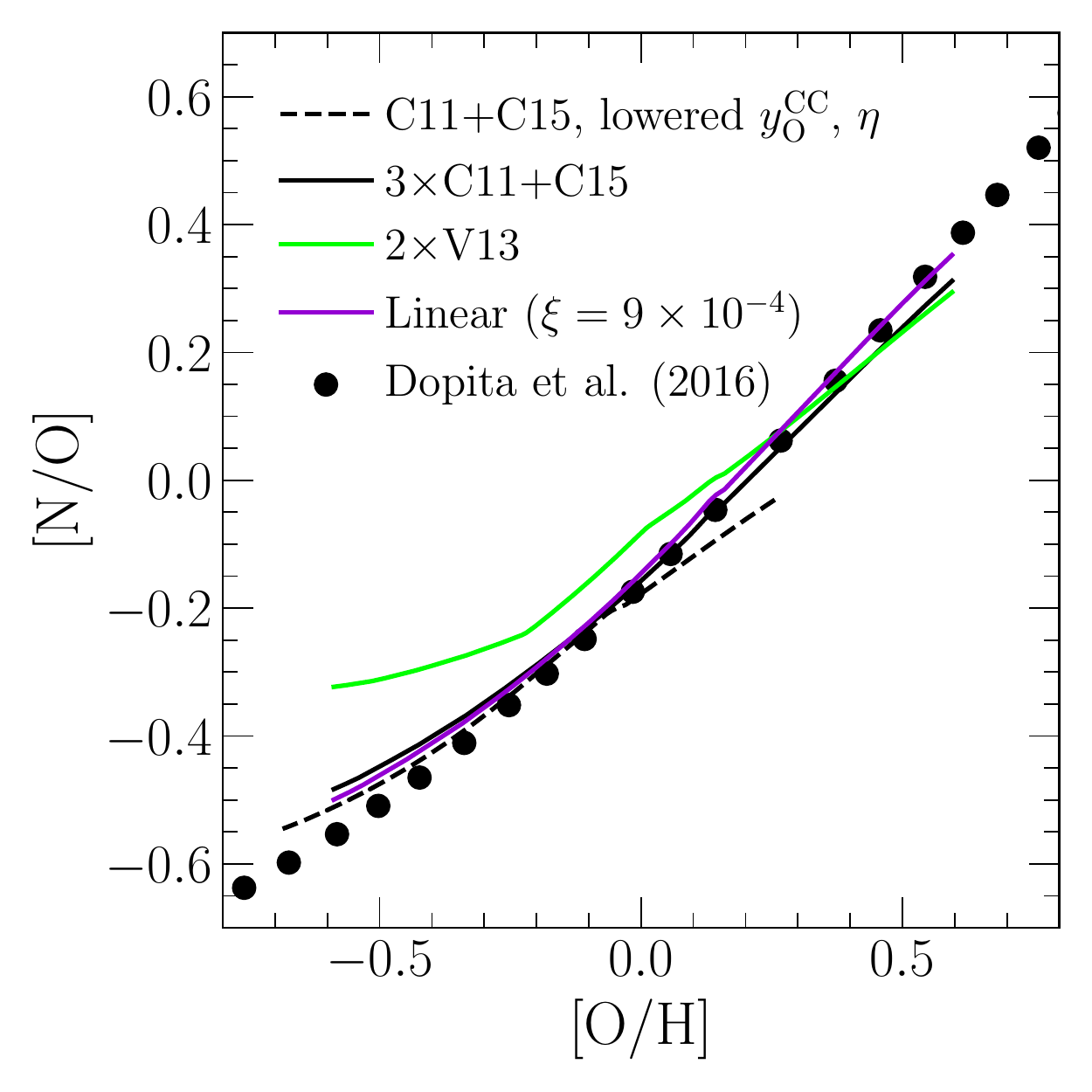}
\includegraphics[scale = 0.45]{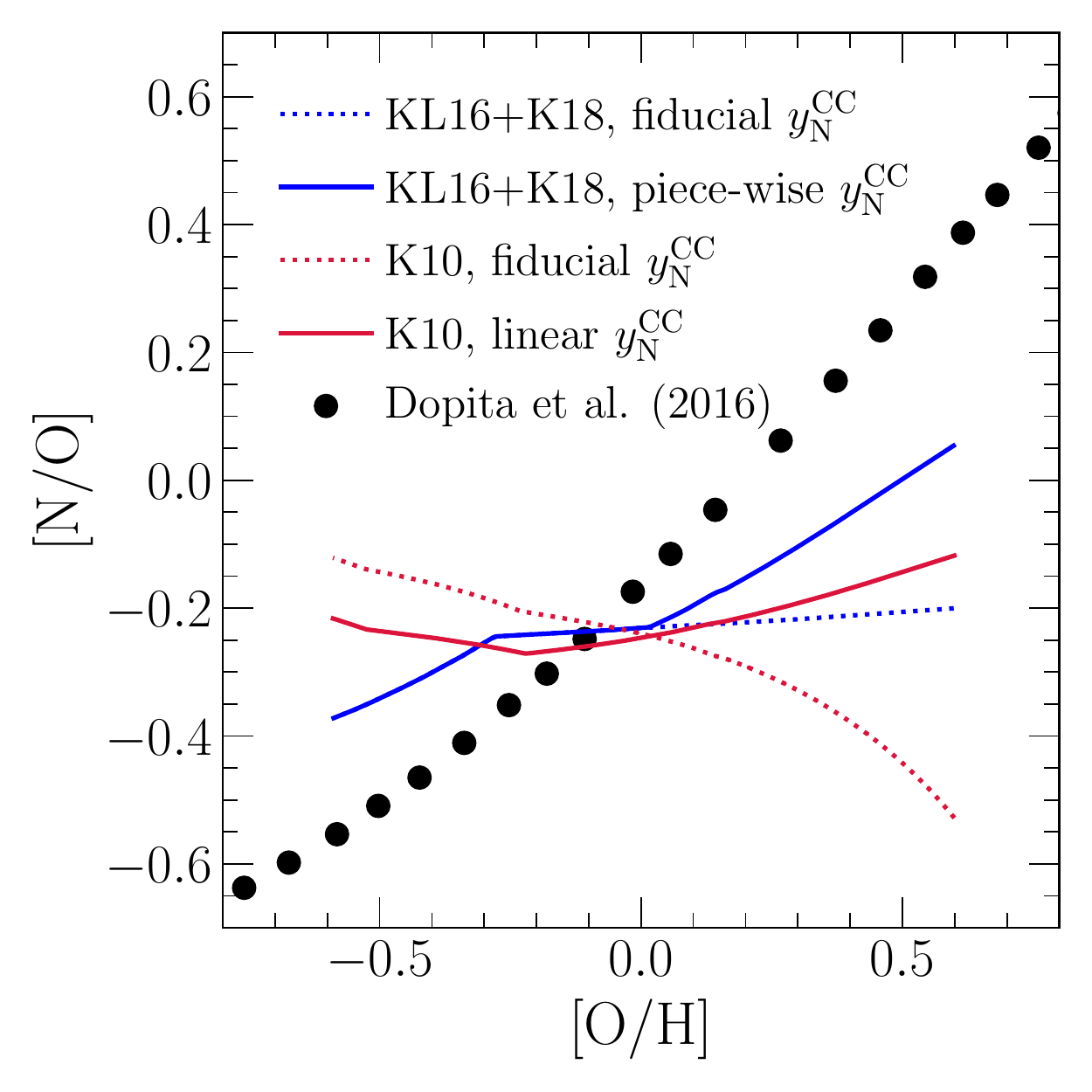}
\caption{
\textbf{Left}: The present-day gas-phase~\ohno~relation predicted by our
model with each of the four AGB star yield tables predict by stellar evolution
models discussed in~\S~\ref{sec:yields:agb}, colour-coded according to the
legend.
We include the~\citet{Dopita2016} empirical relation as the observational
benchmark.
\textbf{Middle}: The same as the left panel, but for a case where we
artificially amplify the~\cristallo~yields by a factor of 3 and
the~\ventura~yields by a factor of 2.
We show our fiducial model using the linear AGB star yields with a slope
of~$\xi = 9\times10^{-4}$ in a solid purple line.
The black dashed line shows a model in which the~\cristallo~yields are
unmodified but SN yields and the outflow mass loading factor are lowered by a
factor of 3 (see discussion in~\S~\ref{sec:results:yields:variations}).
\textbf{Right}: The same as the left panel, but comparing the predictions made
by the~\karakasten~and~\karakas~yields with our fiducial value of
$y_\text{N}^\text{CC}$ (dotted lines, same as left-hand panel) to those with
alternate forms of~$y_\text{N}^\text{CC}$ (solid lines; see
equations~\ref{eq:broken_yncc} and~\ref{eq:linear_yncc} and discussion
in~\S~\ref{sec:results:yields}).
We show all predictions with our post-processing migration prescription (see
discussion in~\S~\ref{sec:multizone}).
}
\label{fig:no_oh_predictions}
\end{figure*}

We use the~\citet{Dopita2016}~\ohno~relation, which is inferred by fitting
local stars and HII regions spanning a wide range of~\oh, as our observational
benchmark.
As seen in Fig.~\ref{fig:no_oh_observed}, the~\citet{Pilyugin2010} ``ONS''
calibration leads to a steeper relation at high~\oh.
Following~\citet{Vincenzo2021}, we adopt the~\citet{Dopita2016} relation
because it agrees well with the trends found for APOGEE disc stars and with
results for MaNGA galaxies~\citep[][see
Fig.~\ref{fig:no_oh_observed}]{Belfiore2017}.
None the less, uncertainties in the observed trends remain, and adopting a
significantly different relation would lead to different conclusions about the
metallicity dependence of N yields.
To make the comparison between different yield models more clear, we neglect
the impact of stellar migration on enrichment rates and make use of the
post-processing migration prescription in this section (see discussion
in~\S~\ref{sec:multizone}).
In each panel of Fig.~\ref{fig:no_oh_predictions}, black points show
the~\citet{Dopita2016} trend.
The purple curve in the middle panel shows the prediction of our fiducial
(linear) yield model, which achieves excellent agreement with this
observational benchmark.
\par
In the left panel of Fig.~\ref{fig:no_oh_predictions}, we compare our model
predictions with each of the AGB star yield tables predicted from stellar
evolution models (see Fig.~\ref{fig:agb_yield_models} and discussion
in~\S~\ref{sec:yields:agb}).
Swapping our linear yields (equation~\ref{eq:linear_yield}) out from the
fiducial model for any of these AGB yields taken from the literature results
in a failure to reproduce the observed~\ohno~relation.
The~\cristallo~and~\ventura~yields are able to reproduce the qualitative trend,
but with an incorrect normalization.
The~\karakasten~and~\karakas~yields, on the other hand, fail to reproduce
the steadily sloped increase of~\no~with~\oh.
The inverse dependence of~\no~with~\oh~predicted by the~\karakasten~AGB star
yields can be understood by the interaction between TDU and HBB (see discussion
in~\S~\ref{sec:yields:agb}).
Both effects are stronger at low metallicity, and since all of
the~\karakasten~stars experiencing HBB also experience TDU (see their table 1),
such a result is unsurprising.
This is also true for the~\karakas~yields, but that model predicts a
relatively flat~\ohno~relation because of updated model inputs regarding
opacity and mass loss and the impact this has on~\Nfourteen~yields (see
discussion in~\S~\ref{sec:yields:agb}).

\subsubsection{Variations in SN Yields and Outflows}
\label{sec:results:yields:variations}

In order to successfully reproduce the observations with
the~\cristallo~and~\ventura~yields, we find that we must artificially amplify
them by factors of~$\sim$3 and~$\sim$2, respectively.
We illustrate the results of these modified yield models and for our fiducial
linear model with a normalization of~$\xi = 9\times10^{-4}$ in the middle panel
of Fig.~\ref{fig:no_oh_predictions}.
Although the~\ventura~model predicts an~\ohno~relation that is slightly
shallower than the~\citet{Dopita2016} data, the predictions are
reasonably within the scatter seen in Fig.~\ref{fig:no_oh_observed}.
\par
As an alternative to amplifying the~\cristallo~and~\ventura~yields, we find
good agreement with the observed relation if we instead lower our SN yields
of N and O.
The equilibrium~\no~ratio is largely determined by the IMF-averaged N and O
yields, so lowering~\ycc{O}~has much the same effect on~\no~as raising the AGB
star N yields.
However, lowering~\ycc{O}~while holding other GCE parameters fixed reduces the
O abundance at each Galactocentric radius, so the model no longer reproduces
the normalization of the observed Galactic~\oh~gradient.
We can mostly repair this second problem by also reducing the outflow mass
loading efficiencies assumed in the model.
For a constant SFR, the equilibrium O mass fraction
is~$Z_\text{O} = \ycc{O}/(1 + \eta - r)$ where~$r \approx 0.4$ is the recycling
factor~\citep{Weinberg2017}.
The black dashed curve in the middle panel of Fig.~\ref{fig:no_oh_predictions}
illustrates a model with the unmodified~\cristallo~yields,~\ycc{O}~lowered by a
factor of 3, and~$\eta$ lowered by a factor of 3 at all radii.
This model exhibits good agreement with the~\citet{Dopita2016} trend because N
and O yields are both below those of the fiducial model by the same factor.
The model reaches solar~\oh~at the solar radius, but it predicts lower~\oh~in
the inner galaxy (i.e. a reduced x-axis span in
Fig.~\ref{fig:no_oh_predictions}) because lowering~$\eta$ by a factor of 3 does
not lower~$1 + \eta - r$ by a factor of 3.
\par
Lowering our SN yields by a factor of~$2 - 3$ is plausible if a substantial
fraction of massive stars collapse directly to black holes as opposed to
exploding as SNe at the ends of their lives.
Our IMF-averaged massive star yields (see discussion in~\S
\ref{sec:yields:ccsne}) are based on a~\citet{Kroupa2001} IMF combined with SN
nucleosynthesis models in which most~$M > 8~\msun$ stars explode as a CCSN
\cite[e.g.][]{Woosley1995, Chieffi2004, Chieffi2013, Limongi2018, Nomoto2013}.
However, many massive stars may collapse to form black holes without a SN
(\citealp*{Pejcha2015, Gerke2015};~\citealp{Sukhbold2016, Ertl2016, Adams2017,
Basinger2021, Neustadt2021}).
With the explosion landscape predicted by their W18 neutrino-driven engine, the
CCSN models of~\citet{Sukhbold2016} predict~\ycc{O}~= 0.0056
\citep{Griffith2021a}, nearly three times lower than our fiducial
value.\footnote{
	This can be calculated with~\vice~using the
	\texttt{vice.yields.ccsne.fractional} function, designed to compute values
	of~\ycc{X} for various elements under a variety of assumptions.
}
Extensive black hole formation would also lower~\ycc{Fe}, and a lower
normalization of~\yia{Fe} may be compatible with observational constraints on
SN Ia rates (see discussion in~\S~\ref{sec:yields}).
\par
Another alternative is to retain high~\ycc{O}~but assume that Galactic winds
preferentially remove SN products relative to AGB products. In particular,
~\citet{Vincenzo2016a} are able to reproduce the~\ohno~relation
in chemical evolution models with the~\ventura~yields by implementing a
differential wind in which outflows remove O but not N from the star forming
gas reservoir.
We find similar results for the~\ventura~yields if we simply add a portion of
the SN products (both CCSN and SN Ia) directly to the outflow, which is
otherwise composed of swept up ambient ISM with the same abundance ratios but
reduced~$\eta$.
If SNe are the sources of outflow-driving winds but AGB stars do not
significantly contribute, it would be reasonable to expect some portion of
the SN ejecta to be swept up by the wind; recent theoretical
\citep{Christensen2018} and observational arguments~\citep*{Chisholm2018}
indeed suggest such a scenario.

\subsubsection{Metallicity-Dependent CCSN Yields of N}
\label{sec:results:yields:yncc}

While the issue for the~\cristallo~and~\ventura~yields is one of normalization,
our models with the~\karakasten~or~\karakas~AGB yields predict a
qualitatively incorrect trend of~\no~with~\oh.
This discrepancy cannot be repaired by changing~\ycc{O}.
However, the N yield is the sum of CCSN and AGB star contributions, so it is
reasonable to ask if plausible changes to the metallicity dependence of the
CCSN yield can compensate.
Motivated by the observed~\no~plateau at low metallicity and by the predictions
of rotating massive star models, we have thus far assumed a
metallicity-independent~$\ycc{N} = 3.6\times10^{-4}$.
However, the non-rotating CCSN models of~\citet{Nomoto2013} suggest
that~\ycc{N}~may increase at super-solar metallicity (see
Fig.~\ref{fig:agb_yield_models}).
We therefore construct the following parametrization for use with
the~\karakas~AGB star yields:
\begin{equation}
\ycc{N} = (3.6\times10^{-4})\max\left(1, \frac{Z}{Z_\odot}\right).
\label{eq:broken_yncc}
\end{equation}
Using this yield combination in our GCE model produces the blue solid curve
in the right panel of Fig.~\ref{fig:no_oh_predictions}.
While agreement with the~\citet{Dopita2016} trend is somewhat improved, this
model is still far from the empirical~\ohno~relation.
Achieving agreement while using the~\karakas~yields would require still higher
CCSN yields at~$Z > Z_\odot$ and somewhat lower CCSN yields at~$Z < Z_\odot$.
\par
Because the~\karakasten~AGB model predicts a high N yield at low metallicity
(Fig.~\ref{fig:ssp}), we combine it with the predicted yields of the
non-rotating CCSN models from~\citet{Limongi2018}, which we approximate as
\begin{equation}
\ycc{N} = (3.6\times10^{-4})\left(\frac{Z}{Z_\odot}\right)
\label{eq:linear_yncc}
\end{equation}
(diagonal dotted line in Fig.~\ref{fig:n_cc_yields}, left).
This combination produces the red solid line in the right panel of
Fig.~\ref{fig:no_oh_predictions}.
Although the discrepancy with the~\citet{Dopita2016} trend is somewhat reduced,
this model still underpredicts~\no~at~$Z > Z_\odot$ and overpredicts~\no~at
$Z < Z_\odot$ by large margins.
\par
We conclude that the metallicity dependence of N yields predicted by
the~\karakas~and, especially,~\karakasten~AGB models are empirically
untenable unless massive star yields are far from theoretical expectations.
The implications for stellar astrophysics are uncertain, but inspection of
Fig.~\ref{fig:agb_yield_models} suggests that the problem of these AGB star
models originates in the coexistence of TDU and HBB over a substantial
mass range (M~$\gtrsim 4~\msun$): in both models, every star that experiences
HBB also experiences TDU.

\subsubsection{Summary}
\label{sec:results:yields:summary}
The gas-phase~\ohno~relation in the Milky Way disc predicted by our GCE model
agrees well with the~\citet{Dopita2016} trend characterizing local stars and
HII regions if we assume a metallicity-independent~$\ycc{N} = 3.6\times10^{-4}$
as suggested by rotating massive star models~\citep{Limongi2018} and the linear
model of AGB yields in equation~\refp{eq:linear_yield}
with~$\xi = 9\times10^{-4}$.
Reproducing the empirical trend with the~\cristallo~(\ventura) AGB yields
requires renormalizing those yields by a factor of 3 (2) or, alternatively,
lowering the assumed CCSN O yield~\ycc{O}~and N yield~\ycc{N}~by the same
factor.
Lowering~\ycc{O}~could be physically justified if a large fraction of massive
stars collapse to black holes without producing CCSNe, and it may be
empirically tenable in a model where~\ycc{Fe},~\yia{Fe}, and outflow mass
loading efficiencies~$\eta$ are all lowered by a similar factor.
The metallicity dependence of AGB star yields predicted by
the~\karakasten~or~\karakas~models, flat or even declining with increasing
metallicity, is difficult to reconcile with the observed~\ohno~trend.

\subsection{Metallicity Dependence vs. Age Dependence}
\label{sec:results:t_z_dep_comp}

\begin{figure}
\centering
\includegraphics[scale = 0.6]{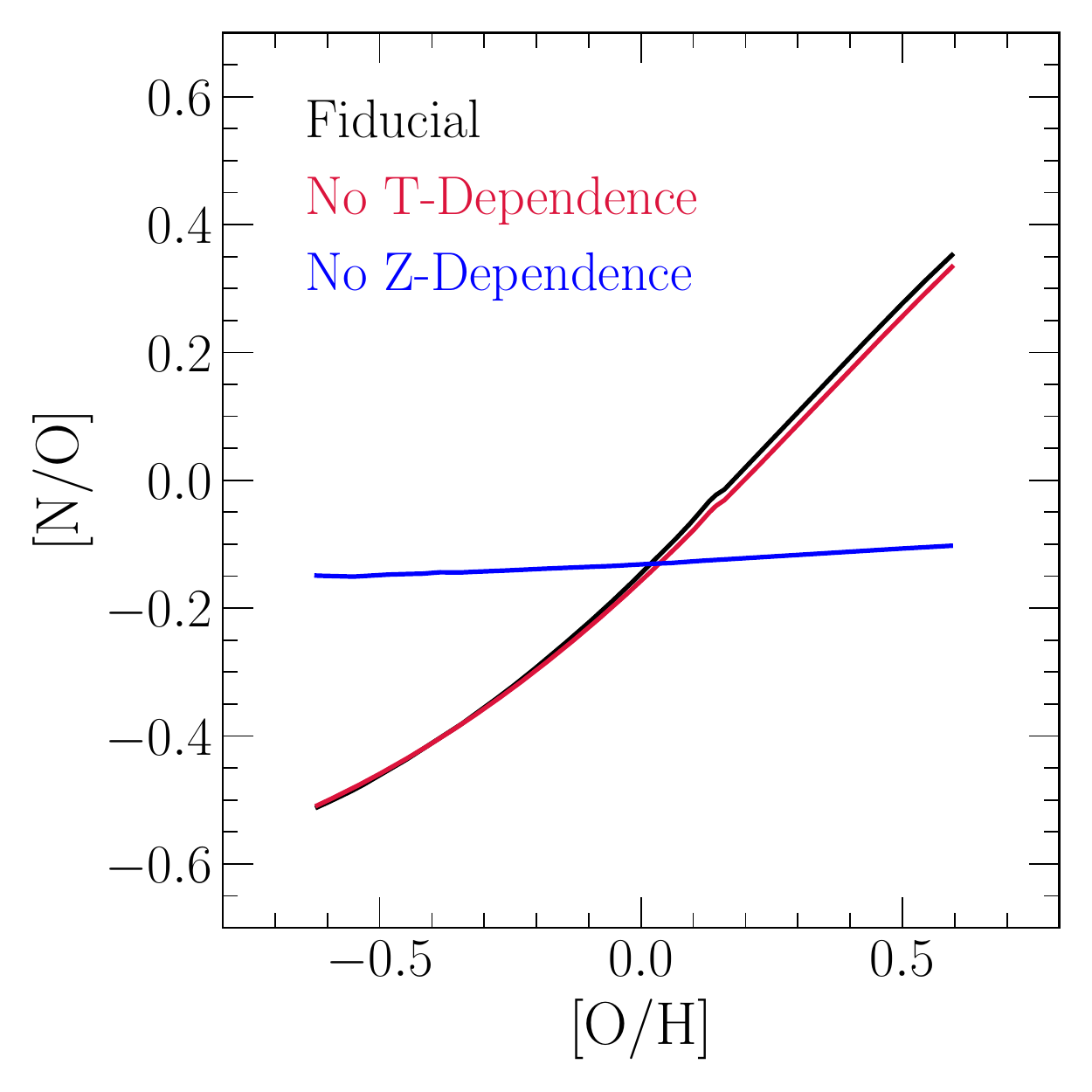}
\caption{
A comparison between our fiducial model with post-processing migration (black)
and variations with the time dependence (red) and metallicity dependence 
removed (blue).
To remove the time dependence, we pre-compute the AGB star yields of N from
13.2 Gyr old stellar populations as a function of metallicity as in the right
panel of Fig.~\ref{fig:ssp}, then incorporate this into the prompt CCSN yields
and set the delayed AGB star contribution to zero.
To remove the metallicity dependence, we evaluate the yields at our assumed
solar metallicity of~$Z_\odot = 0.014$ at all timesteps.
}
\label{fig:t_z_dep_comp}
\end{figure}

Despite predicting a different mass dependence for~\yagb{N}~(see Fig.
\ref{fig:agb_yield_models}), the renormalized~\cristallo~and~\ventura~yields
both reproduce the~\ohno~relation reasonably well.
This result suggests that the metallicity dependence plays a much more
important role than the DTD in establishing this correlation.
To investigate this point further, we consider two variants of our fiducial
model: one with the dependence on stellar age (or, equivalently, stellar mass)
removed from the enrichment rate calculations, and the other with the
metallicity dependence removed.
We return to our linear AGB yield model with~$\xi = 9\times10^{-4}$, and
to make this comparison more straightforward we use the post-processing
migration model (see discussion in~\S~\ref{sec:multizone}).
\par
To remove the age dependence, we simply eject the AGB star yields alongside
the CCSN yield instantaneously after a single stellar population forms.
We pre-compute the N yields from all AGB stars associated with a 13.2 Gyr old
stellar population as a function of progenitor metallicity in a similar fashion
as in the right panel of Fig.~\ref{fig:ssp}.
Since~\vice~works from IMF-averaged CCSN yields assumed to be injected
instantaneously following a single stellar population's formation (see
discussion in~\S~\ref{sec:yields:ccsne}), we make use of the software's
capability to let the user specify functional forms for nucleosynthetic yields
and simply add this N yield to~\ycc{N}~and set~\yagb{N}~to zero.
In this model,~\ycc{N}~inherits a metallicity dependence from the AGB star
yields and has the exact shape of the purple curve in the right hand panel of
Fig.~\ref{fig:ssp}.
To remove the metallicity dependence, the procedure is much simpler: we simply
evaluate~\yagb{N}~at our assumed solar metallicity of~$Z_\odot = 0.014$ at all
timesteps regardless of that which is predicted for a stellar population.
In this variation, AGB star production still occurs on a DTD inherited from
the stellar mass-lifetime relation~\citep{Larson1974} and the mass dependence
of the linear yield model.
\par
We illustrate these predictions in Fig.~\ref{fig:t_z_dep_comp}.
The~\ohno~relation from the model with no age dependence is nearly identical to
the prediction found in our fiducial model, while the prediction with no
metallicity dependence is considerably different.
This result is rather unsurprising given the short characteristic timescales of
N production ($\sim$250 Myr, see the middle panel of Fig.~\ref{fig:ssp}).
Mathematically, there is little difference in the enrichment rates if all of a
stellar population's N is produced immediately as opposed to from a prompt,
sharply declining DTD.
The metallicity dependence, however, is paramount to the~\ohno~relation, which
is expected given the results in Fig.~\ref{fig:agb_yield_models} and consistent
with previous arguments that the increase in~\no~at high~\oh~is a consequence
of secondary N production~\citep{VilaCostas1993, vanZee1998, Henry1999,
PerezMontero2009, Berg2012, Pilyugin2012, Andrews2013, HaydenPawson2021}.
Fig.~\ref{fig:t_z_dep_comp} implies that the gas-phase~\ohno~relation offers
little if any constraining power over the mass dependence of N yields from
AGB stars.

\subsection{Comparison to Stellar Abundances in the Milky Way Disc}
\label{sec:results:vincenzo_comp}

\begin{figure}
\centering
\includegraphics[scale = 0.6]{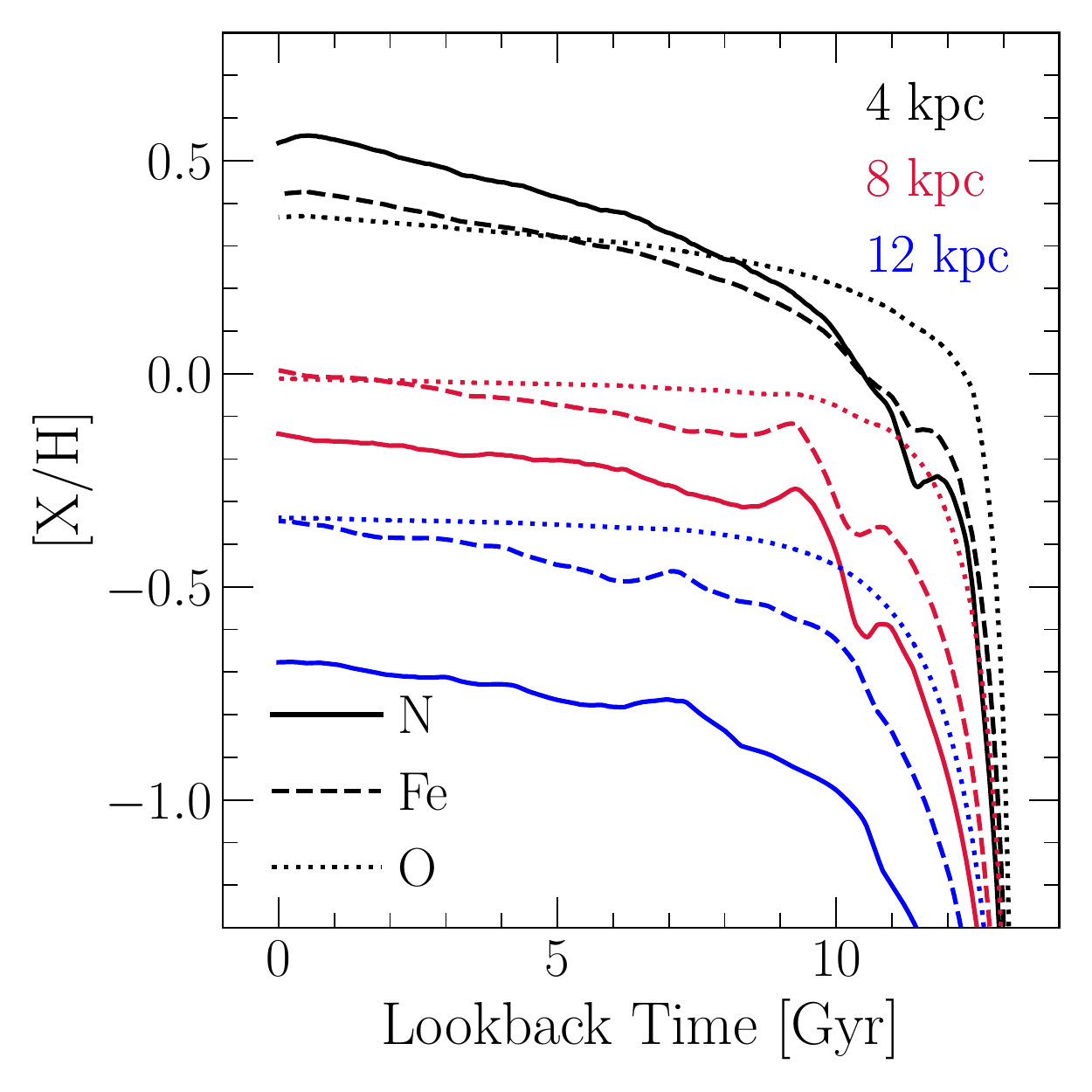}
\caption{
\nh~(solid),~\feh~(dashed), and~\oh~(dotted) in the gas-phase as a function of
lookback time in the fiducial model at~$\rgal = 4$ (black), 8 (red), and 12 kpc
(blue).
}
\label{fig:nh_feh_vs_lookback}
\end{figure}

\begin{figure*}
\centering
\includegraphics[scale = 0.6]{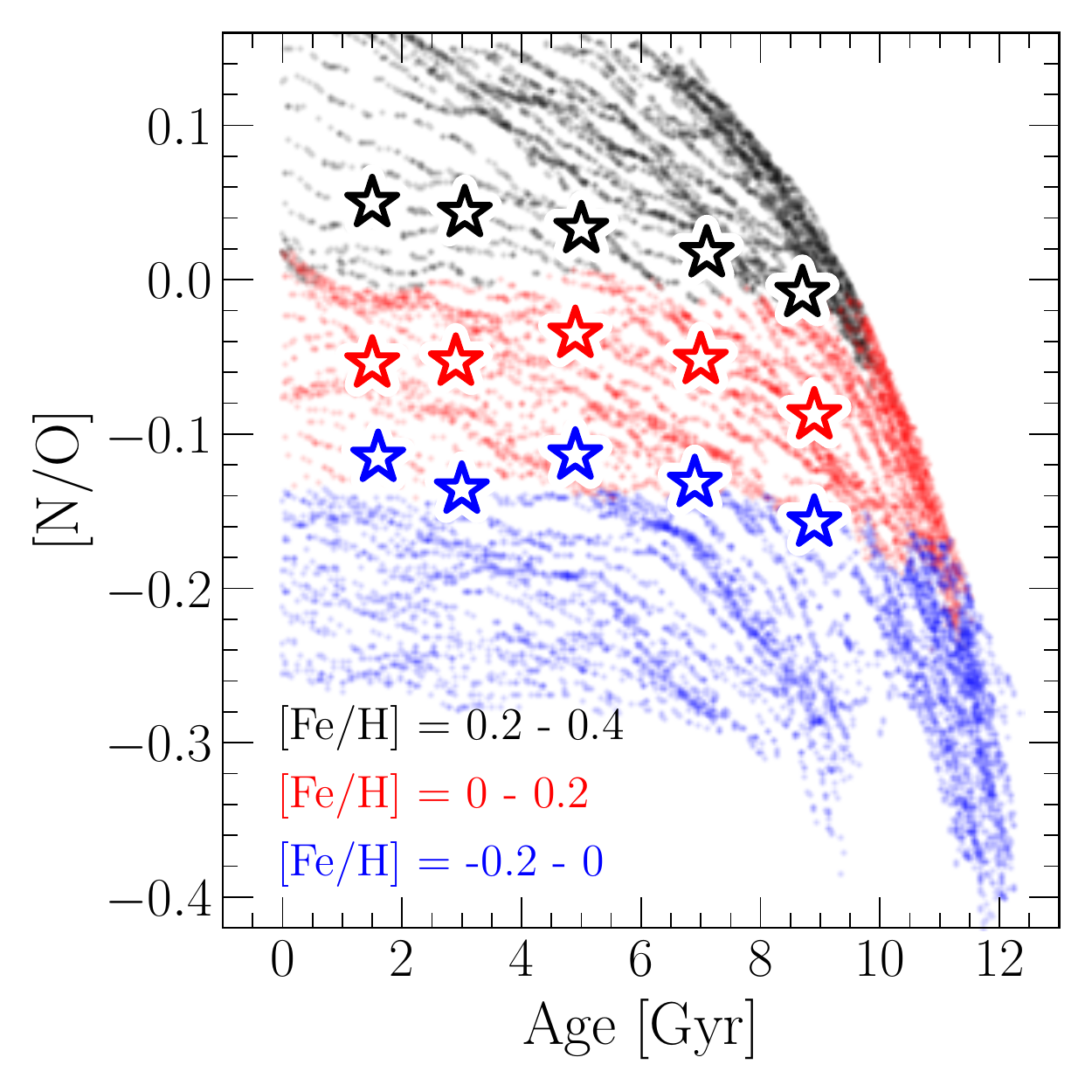}
\includegraphics[scale = 0.6]{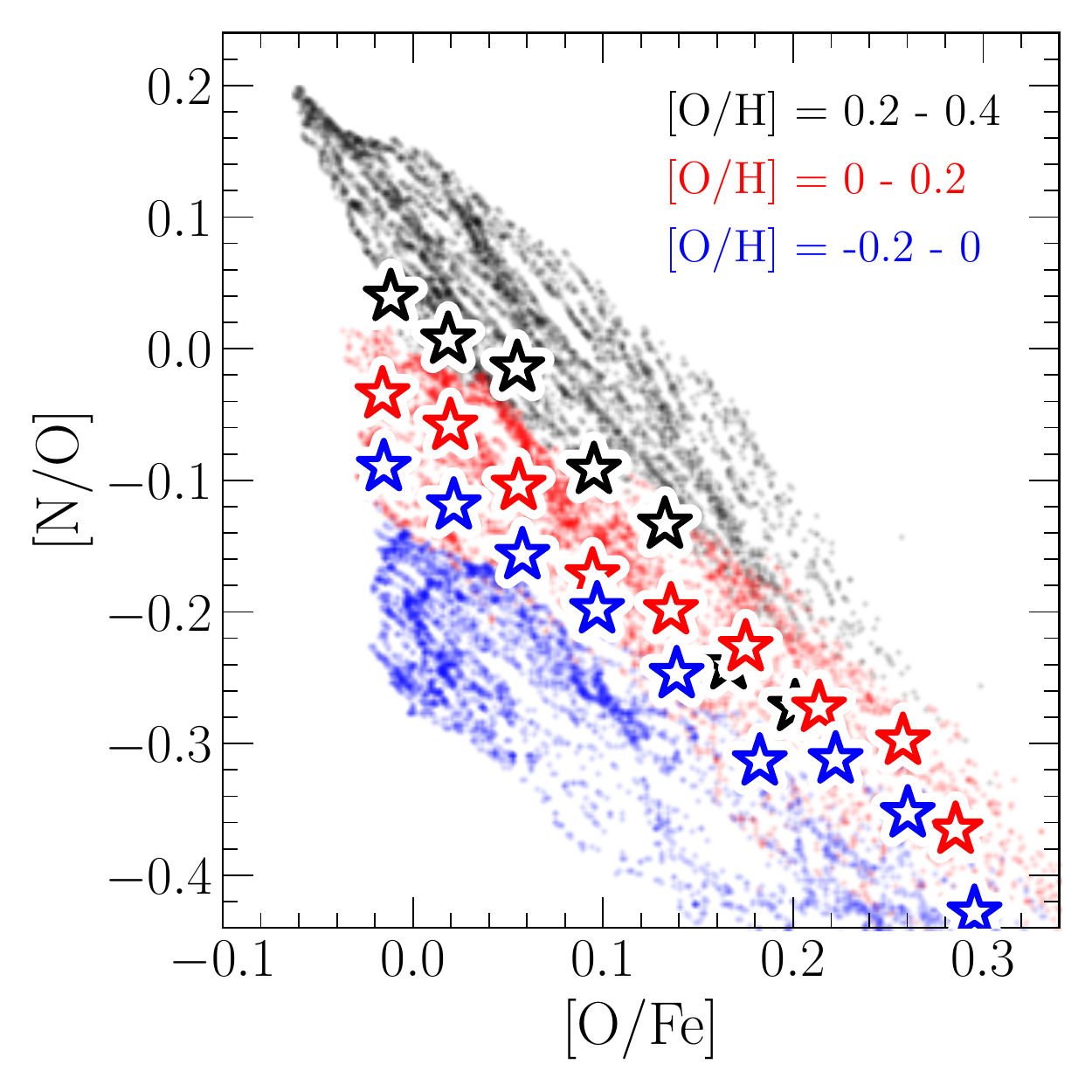}
\caption{
\textbf{Left}:~\no~as a function of stellar age for 5000 stars randomly
sampled from our model stellar populations in three bins of~\feh~(coloured 
points).
Stars quantify the median trend in~\no~with age using N abundances corrected
for internal mixing processes reported by~\citet{Vincenzo2021} in the same
bins of~\feh.
\textbf{Right}: The same as the left panel, but instead showing~\no~as a
function of~\ofe~in bins of~\oh.
}
\label{fig:vincenzo_comp}
\end{figure*}

Although N abundances are typically measured in the gas-phase in external
galaxies, APOGEE~\citep{Majewski2017} has measured N abundances in large
stellar samples spanning many regions of the Milky Way.
By additionally making use of these data, we can investigate trends with
stellar age and~\ofe~at fixed metallicity.
Before comparing the predictions of GCE models to N abundances derived from
spectra of red giant samples such as APOGEE, it is essential to adjust the
measurements for internal processes known to alter the surface compositions of
stars because GCE models predict the birth abundances.
During the main sequence lifetime of M~$\gtrsim 1.3~\msun$ stars, the CNO cycle
processes much of the C and O nuclei in the core into~\Nfourteen.
When the star evolves off the main sequence, this N-rich material is mixed with
the outer convective layers, increasing the N abundance in the
photosphere~\citep{Gilroy1989, Korn2007, Lind2008, Souto2018, Souto2019}.
Using~\texttt{MESA} stellar evolution models~\citep{Paxton2011, Paxton2013,
Paxton2015, Paxton2018} with standard mixing prescriptions,~\citet{Vincenzo2021}
developed a recipe to approximate the birth abundances of C, N, and O and
applied it to the sample of APOGEE/Kepler red giants with asteroseismic mass
measurements from~\citet{Miglio2021}.
They found good agreement between the mean trend of~\no~with~\oh~for APOGEE
disc stars and the~\citet{Dopita2016} trend.
Since our fiducial model reproduces the~\citet{Dopita2016} trend over most of
its history (Fig.~\ref{fig:no_oh_timeevol}), it should also reproduce
the~\ohno~relation for APOGEE disc stars.
\par
Our models predict a correlation between N and Fe abundances in
the gas-phase which turns out to be important to understanding how the model
predictions compare to stellar abundances.
In Fig.~\ref{fig:nh_feh_vs_lookback}, we plot the evolution of~\nh,~\oh,
and~\feh~in the ISM at~$\rgal = 4$, 8, and 12 kpc in our fiducial model with
linear AGB yields (equation~\ref{eq:linear_yield}) and the diffusion migration
prescription (see discussion in~\S~\ref{sec:multizone}).
\nh~is more correlated with~\feh~than~\oh~at all radii, and the relation
persists up to lookback times of~$\sim$10 Gyr.
This arises in part because N and Fe are both produced in significant
quantities by delayed enrichment sources while O is produced almost entirely on
short timescales by CCSNe (see discussion in~\S~\ref{sec:yields}).
Although the production timescale of N from single stellar populations is
short (see discussion in~\S~\ref{sec:yields:imf_agb}), metallicity dependent
yields require more abundant species such as O to be produced and reach an
equilibrium before N yields stabilize.
When many stellar populations are present, the bulk of the N production will
thus always follow the bulk production of more abundant species; this is 
qualitatively similar to what~\citet{Johnson2020} found regarding the
production timescales of Sr and Fe.
As a consequence of both its slight delay and its metallicity-dependent
yields, N reaches its equilibrium abundance on timescales similar to Fe rather
than O.
Due to the prompt and metallicity independent nature of O enrichment,~\oh~is
near equilibrium as far back as~$\sim$10 Gyr ago while~\nh~and~\feh~are not.
\par
Combining the~\citet{Vincenzo2021}~\no~ratios with the APOGEE stellar ages
taken from~\citet{Miglio2021}, we illustrate the~\no-age relation in bins
of~\feh~as predicted by our model and measured by APOGEE in the left panel of
Fig.~\ref{fig:vincenzo_comp}.
In good agreement with the observational measurements, the model predicts
the~\no-age relation to be relatively flat in bins of~\feh.
This arises as a consequence of the N-Fe correlation and the fast approach to
equilibrium in~\oh~as discussed above (see Fig.~\ref{fig:nh_feh_vs_lookback}).
A bin in~\feh~approximately corresponds to a bin in~\nh, and by extension a bin
in~\no~as well since~\oh~is nearly constant at fixed radius up to~$\sim$10 Gyr
ago.
\par
In the right panel of Fig.~\ref{fig:vincenzo_comp}, we compare our model
predictions to the~\no-\ofe~relation at fixed~\oh~reported by
\citet{Vincenzo2021}.
The model correctly predicts a significant inverse relationship
between~\no~and~\ofe.
This is again a consequence of the N-Fe correlation demonstrated in
Fig.~\ref{fig:nh_feh_vs_lookback}:~\nh~increases with~\feh, so at
fixed~\oh,~\no~increases as~\ofe~decreases.
This is another important success of our model.
\citet{Vincenzo2021} demonstrate that high [$\alpha$/Fe] and low [$\alpha$/Fe]
disc populations show a dichotomy in~\no.
The~\citet{Johnson2021} GCE model produces a broad but continuous [$\alpha$/Fe]
distribution rather than the bimodal distribution found in previous works
\citep[e.g.][]{Hayden2015, Vincenzo2021b}, but Fig.~\ref{fig:vincenzo_comp}
suggests that a model tuned to produce the [$\alpha$/Fe] bimodality would also
produce a dichotomy in~\no.
\par
Quantitatively, our model slightly underpredicts~\no~in the lower metallicity
bins in both panels of Fig.~\ref{fig:vincenzo_comp}.
In general, our model occupies a noticeably wider range in~\no~than do the
\citet{Vincenzo2021} measurements at all ages and all~\ofe.
This could be a sign that the AGB star yields of N in our fiducial model scale
slightly too strongly with the total metallicity~$Z$.
Since our fiducial model assumes an exactly linear scaling of the N yield
with~$Z$ (see equation~\ref{eq:linear_yield}), this suggests that perhaps a
slightly sub-linear scaling would be more accurate, but only barely because
the discrepancies in Fig.~\ref{fig:vincenzo_comp} are at the~$\sim$0.1 dex
level.
\par
Although we demonstrate in~\S~\ref{sec:results:t_z_dep_comp} that the
metallicity dependence of the yield plays the strongest role in establishing
the~\ohno~relation, the DTD plays a significant role in shaping the stellar
abundances.
In the model in which AGB nucleosynthetic yields are injected instantaneously
along with CCSN products, the N-Fe correlation described above is no longer
present.
Instead, N approaches equilibrium on much faster timescales, resulting in it
being much more correlated with O than Fe, and the resulting~\no-age relation
is positively sloped at fixed~\feh.
This suggests that the DTD may play a minimal role in establishing gas-phase
abundances but is important in shaping stellar abundances.
We find additional discrepancies if we instead attribute the delayed N
production to a metallicity-dependent SN Ia yield with the same~$t^{-1.1}$ DTD
that we adopt for Fe.
The agreement in Fig.~\ref{fig:vincenzo_comp} suggests that our fiducial model
has a fairly accurate separation of CCSN and AGB contributions and that the
$\sim$250 Myr characteristic delay for AGB enrichment predicted by this model
is approximately correct -- instantaneous N enrichment is too fast and
$t^{-1.1}$ enrichment is too slow to match the observations.

\subsection{The Sources of Scatter in the~\ohno~Relation}
\label{sec:results:schaefer_comp}

\begin{figure*}
\centering
\includegraphics[scale = 0.44]{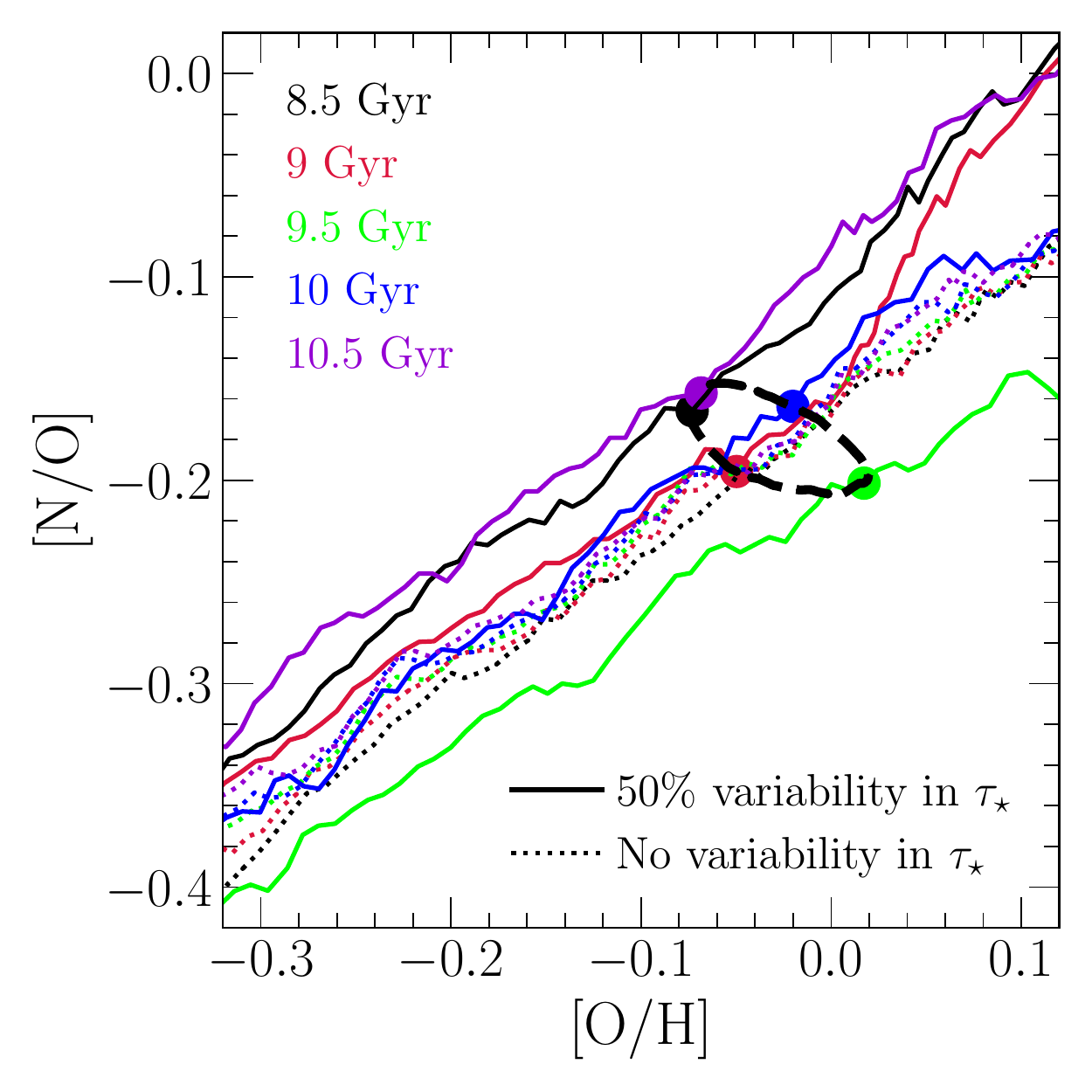}
\includegraphics[scale = 0.47]{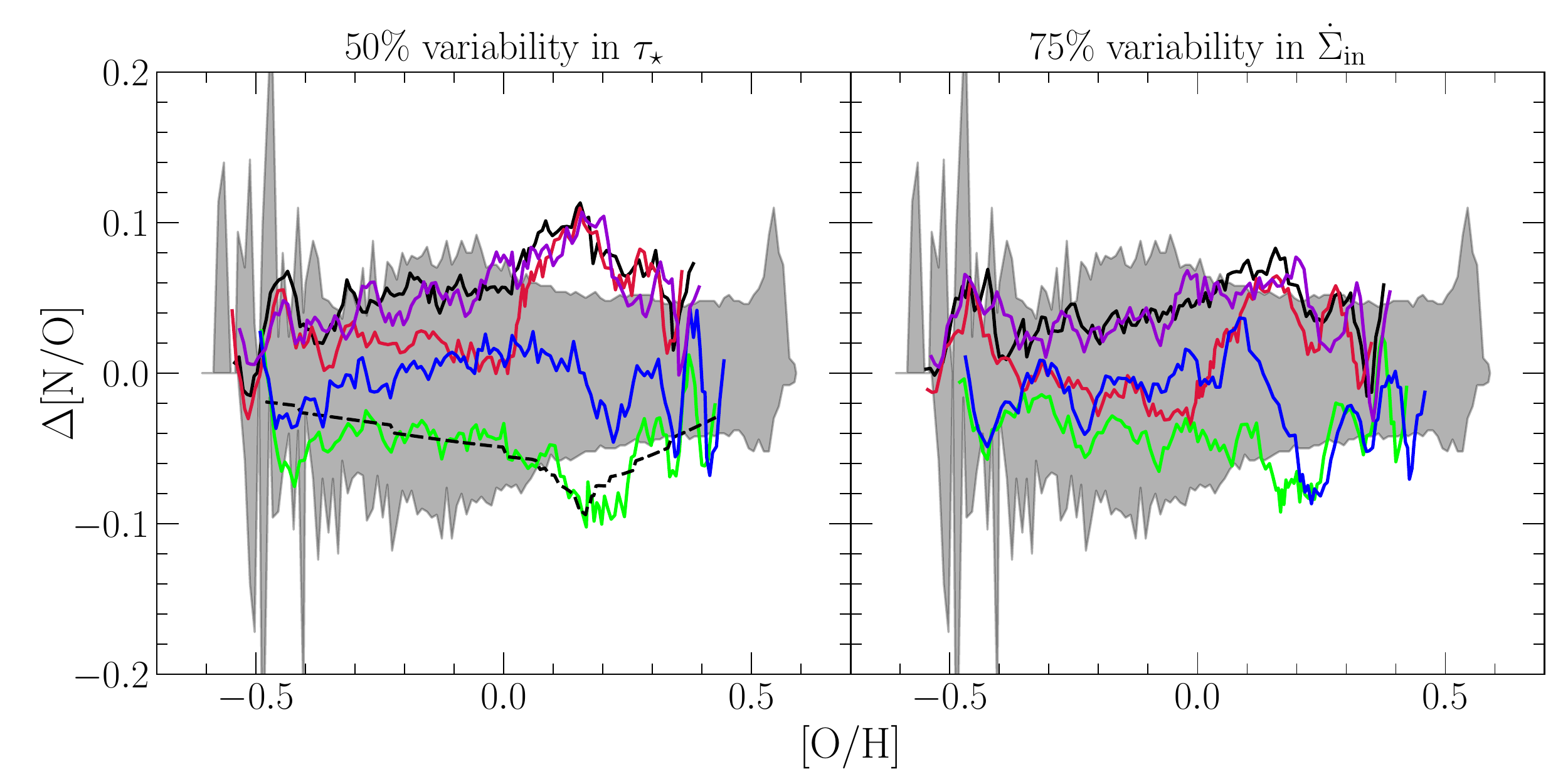}
\caption{
\textbf{Left}: One cycle of oscillations in the~\ohno~relation at
high~\oh~induced by sinusoidal variability in~$\tau_\star$ with an amplitude of
0.5 (solid coloured lines; see equation~\ref{eq:sfe_var}).
Dotted lines show the~\ohno~relation at the same five snapshots in the fiducial
model with no variability in~$\tau_\star$.
We use the diffusion migration prescription in both cases (see discussion
in~\S~\ref{sec:multizone}).
The black dashed line shows the time evolution of the abundances at~$\rgal = 8$
kpc, the approximate Galactocentric radius of the sun, with the times of each
of the five snapshots marked by a coloured point.
\textbf{Middle and Right}: For the same five snapshots in the left hand panel,
the deviation in~\no~at fixed~\oh~relative to the fiducial model for the case
with sinusoidal variability in~$\tau_\star$ at an amplitude of 0.5 (middle; see
equation~\ref{eq:sfe_var})
and with sinusoidal variability in~$\dot{\Sigma}_\star$ at an amplitude of 0.75
(right; see equation~\ref{eq:ifr_var}).
The shaded regions in both panels quantify the width of the [N/O] distribution
in~$10^{10.5} - 10^{11}~\msun$ galaxies in MaNGA taken from
\citet{Schaefer2020}.
In bins of~\oh, we place the median~\no~at~$\Delta\no = 0$, and the lower
(upper) envelope denotes the 16th (84th) percentile of the~\no~distribution.
The black dashed line in the middle panel denotes the same quantity as the
corresponding solid green line but computed from the post-processing migration
prescription, which neglects the impact of stellar migration in computing
enrichment rates (see discussion in~\S~\ref{sec:multizone}).
}
\label{fig:schaefer_comp}
\end{figure*}

\citet{Schaefer2020} demonstrate that intrinsic scatter in the
gas-phase~\ohno~relation is correlated with variations in the local SFE.
This is expected from one-zone GCE models~\citep[e.g.][]{Molla2006,
Vincenzo2016a}.
Although we have demonstrated in~\S~\ref{sec:results:fiducial} that the impact
of stellar migration on enrichment rates is small, it could none the less
contribute additional scatter in the observed~\ohno~relation.
Our models, taking into account the effects of migration on the
enrichment rates while allowing full control over the SFH and the SFE
through~\vice, are an ideal tool with which to address this question.
\par
To this end, we construct two variants of our fiducial model
with linear AGB yields (equation~\ref{eq:linear_yield}) and the
diffusion migration prescription (see discussion in~\S~\ref{sec:multizone}).
While the fiducial model specifies the SFH~\textit{a priori} and
lets~\vice~compute the infall history~$\dot{\Sigma}_\text{in}$, here we specify
the infall history as a function of radius and time.
As we will demonstrate below, the effects of dilution play an important role in
driving variations in the~\ohno~plane in these variants, and by specifying the
infall history we have more control over the amount of dilution.
In a similar fashion as in our fiducial model, we normalize
$\dot{\Sigma}_\text{in}$ such that a stellar mass consistent with that reported
by~\citet{Licquia2015} arises from the simulation.
All other evolutionary parameters are the same as described
in~\S~\ref{sec:multizone}.
\par
In the first variant, the SFE exhibits 50\% sinusoidal oscillations with time
over 2 Gyr periods:
\begin{equation}
\tau_\star(\rgal, t) = \tau_{\star,\text{J21}}(\rgal, t)
\left(1 + 0.5\sin\left(\frac{2\pi t}{2~\text{Gyr}}\right)\right).
\label{eq:sfe_var}
\end{equation}
The infall rate is constant in each ring with a value determined by normalizing
to the present day stellar mass and stellar surface density gradient of the
Milky Way.
Our choice of a 50\% amplitude is comparable to the observationally derived
scatter in molecular gas depletion times according to multiple measurement
methods (see figs. 4 and 5 of~\citealp{Tacconi2018} and references therein).
Furthermore, variations in the SFE are of similar magnitude in~\hsim, the
galaxy from which our model's migration history is drawn.
In the second variant, the SFE is constant and the infall rate oscillates with
a 75\% amplitude about its value in the first variant:
\begin{equation}
\dot{\Sigma}_\text{in}(\rgal, t) = \langle\dot{\Sigma}_\text{in}\rangle
\left(1 + 0.75\sin\left(\frac{2\pi t}{2~\text{Gyr}}\right)\right).
\label{eq:ifr_var}
\end{equation}
The amplitude of 75\% is chosen such that the ensuing variability in the SFR is
of similar magnitude between the two models ($\sim$40\%).
We additionally run a model in which neither the accretion rate nor the SFE
oscillate, replacing the fiducial model used in previous sections with this
constant infall model.
Otherwise, the evolutionary differences beyond simple oscillations complicate
the comparison.
\par
These variant models characterize evolutionary pathways in which star formation
is more episodic than previously explored.
In a real galaxy, variability in the SFE and the SFR is likely non-sinusoidal
and not with constant amplitude.
A sample of galaxies will have different amplitudes and be seen at different
phases in their variability, and the impact of this on their N and O abundances
will present as intrinsic scatter in a sufficiently large sample.
By comparing models with and without reasonable amounts of variability in these
quantities while taking into account radial migration, we can assess which
quantities impact abundances more strongly and are thus the more likely causes
of intrinsic scatter in the observed~\ohno~relation.
\par
In the left panel of Fig.~\ref{fig:schaefer_comp}, we plot the predicted
gas-phase~\ohno~relation for five snapshots covering one cycle of fluctuations
induced by variability in~$\tau_\star$ according to equation~\refp{eq:sfe_var}.
This model predicts a~$\sim$0.1 dex dynamic range in~\no~at fixed~\oh, whereas
the constant model with no variability in~$\tau_\star$ predicts the relation
to be quite steady over this time interval.
This suggests that stellar migration, present in both the constant model and
this oscillatory variant, does not induce significant variability in
the~\ohno~plane; however, we demonstrate below that its effects are none the
less non-negligible.
The minimal impact of stellar migration traces back to the timescales of N
production from single stellar populations (see Fig.~\ref{fig:ssp} and
discussion in~\S~\ref{sec:yields:imf_agb}): with most N production occurring
within~$\sim$250 Myr of a stellar population's formation, most stars will not
migrate far from their birth radius by the time they produce most of their N,
and the resulting impact on abundances is small.
\par
The behavior in the~\ohno~plane predicted by the oscillatory SFE variant
is driven by the tug-of-war between dilution and re-enrichment
associated with oscillations in~$\tau_\star$.
When star formation quickens, O production increases in proportion.
The ISM abundance and consequently the N yields increase as well.
Because of the slight but none the less finite delay-time of its production by
AGB stars, the N enrichment rate lags slightly behind O.
\no~therefore decreases, and the ISM moves down and to the right in
the~\ohno~plane.
When star formation eventually slows, O production again follows suit.
The N enrichment rate, as before, lags slightly behind, and~\no~increases; the
ISM therefore moves up and to the left in the~\ohno~plane.
The result is an anti-clockwise loop, which we illustrate for the solar circle
with a black dashed line in the left panel of Fig.~\ref{fig:schaefer_comp}.
The effect is generally larger in~\oh~than in~\no~($\sim$0.1 dex versus
$\sim$0.05 dex in this example) because dilution affects
both~\oh~and~\nh~similarly.
In the model with oscillations in~$\dot{\Sigma}_\text{in}$, we find that
qualitatively similar processes drive the evolution in abundances, but there
are interesting differences in detail which we discuss below in the context of
scatter in the observed trend.
\par
In the middle and right panels of Fig.~\ref{fig:schaefer_comp}, we plot the
scatter in the gas-phase~\ohno~relation inferred observationally by
\citet{Schaefer2020}.
Using data from the MaNGA IFU survey~\citep{Bundy2015}, they measure N and O
abundances in 709,541 spaxels across 6,507 unique galaxies spanning
$10^9 - 10^{11}~\msun$ in stellar mass.
Since our model is appropriate for Milky Way mass galaxies, we focus our
comparison on the~$M_\star = 10^{10.5} - 10^{11}~\msun$ mass range
\citep{Licquia2015}, which cuts our sample sample to 197,787 individual N and O
measurements from the MaNGA IFU spaxels.
In narrow bins of~\oh, we then compute the 16th, 50th, and 84th percentiles of
the~\no~distribution.
Placing the median~\no~at~$\Delta\no = 0$, the shaded regions above and below
0 in Fig.~\ref{fig:schaefer_comp} denote the difference between 16th and 84th
percentiles of the distribution in each~\oh~bin.
\par
We compare both of our oscillatory variants to the width of the~\no~distribution
by over-plotting the difference in~\no~at fixed~\oh~between our oscillatory
models and their constant counterpart (i.e. the vertical offset between the
solid and dotted lines in the left panel, and the equivalent thereof for the
oscillatory~$\dot{\Sigma}_\text{in}$ model).
Both models produce offsets in~\no~at fixed~\oh~which, as discussed above,
arise as consequences of dilution, and the offsets are generally consistent
with the width of the relation derived observationally by~\citet{Schaefer2020}.
This supports their argument that variations in the local SFE can drive
intrinsic scatter in the~\ohno~relation, but the effects of stellar migration
are still non-negligible.
We demonstrate this by comparing the green solid line in the middle panel for
the~$t = 9.5$ Gyr snapshot to the black dashed line denoting the same quantity
with post-processing migration (see discussion in~\S~\ref{sec:multizone}).
While some of the features in~$\Delta\no$ as a function of~\oh~can be
attributed to the difference in GCE parameters, stellar migration
affects~\no~ratios with an amplitude of~$\sim$0.05 dex (see also
Fig.~\ref{fig:no_oh_timeevol} and discussion in~\S~\ref{sec:results:fiducial}).
Although this is smaller than the impact of oscillations in either~$\tau_\star$
or~$\dot{\Sigma}_\text{in}$ ($\sim$0.1 dex), it is none the less significant
compared to the width of the~\citet{Schaefer2020} distributions, also at
the~$\lesssim$0.1 dex level; this suggests that stellar migration is a
subdominant but non-negligible source of scatter.
\par
In general, variability in~$\tau_\star$ impacts abundances more strongly than
variability in~$\dot{\Sigma}_\text{in}$.
Fig.~\ref{fig:schaefer_comp} shows similar changes in~\no~at fixed~\oh~in both
of our oscillatory variants, but it requires an amplitude of 75\% in accretion
rates to achieve the same~$\Delta\no$ as an amplitude of 50\% in the SFE.
This weaker impact arises out of an abundance response that is much
more~\textit{along} the~\ohno~relation rather than~\textit{against} it as in the
oscillatory~$\tau_\star$ model (see the black dashed line in the left panel of
Fig.~\ref{fig:schaefer_comp}).
Both~\oh~and~\nh~vary with larger amplitudes in the variable infall model due to
episodes of enhanced and suppressed accretion, but the effects of dilution
on~\nh~are amplified by the combination with metallicity-dependent yields.
As a result,~\nh~varies with a larger amplitude than~\oh, whereas the opposite
is the case in the oscillatory~$\tau_\star$ model.
Consequently,~\no~increases rather than decreases with increasing~\oh, and
changes in~\no~at fixed~\oh~are smaller.
In the context of the observational results~\citep{Schaefer2020}, this suggests
that different~\no~ratios at fixed~\oh~are less likely to reflect changes in
the accretion rate and more likely to reflect variations in the internal
properties of the star forming ISM, such as the thermal state or the pressure,
quantities which here get folded into~$\tau_\star$.

\subsection{The~\ohno~Relation as an Equilibrium Phenomenon}
\label{sec:results:ohno_equilibrium}

As shown by~\citet{Weinberg2017}, under generic conditions the abundances of a
one-zone GCE model with continuing gas infall evolve to an equilibrium, in
which new metal production is balanced by dilution and by the loss of ISM
metals to star formation and outflows.
Fig.~\ref{fig:no_oh_timeevol} shows that the~\no~ratio at a given radius in our
multi-zone GCE model approaches an equilibrium after~$t \approx 5$ Gyr.
Fig.~\ref{fig:t_z_dep_comp} further shows that the~\ohno~relation that emerges
in our model is driven by the metallicity dependence of the N yield, with the
time-delay of AGB enrichment having minimal impact.
This indicates that -- for the purposes of computing the equilibrium abundance
of N -- the AGB star DTD can be neglected, assuming instantaneous production as
in~\S~\ref{sec:results:t_z_dep_comp}.
This allows analytic solutions to the~\no~ratio that will arise at a
given equilibrium~\oh.
\par
For an exponential SFH,~$\dot{M}_\star \propto e^{-t/\tau_\text{SFH}}$, with
instantaneous enrichment and recycling of an element X with IMF-averaged
yield~$y_\text{X}$, the equilibrium ISM mass fraction is
\begin{equation}
Z_\text{eq,X} = \frac{y_\text{X}}{1 + \eta - r - \tau_\star/\tau_\text{SFH}},
\label{eq:zeq}
\end{equation} 
where (as before)~$\eta = \dot{M}_\text{out} / \dot{M}_\star$, $r \approx 0.4$
is the recycling fraction, and~$\tau_\star = M_\text{gas} / \dot{M}_\star$ is
the SFE timescale (see section~\ref{sec:multizone}).
When an element's characteristic enrichment delay time is~$\ll \tau_\text{SFH}$,
the correction to equation~\refp{eq:zeq} is very small~\citep{Weinberg2017},
which alongside Fig.~\ref{fig:t_z_dep_comp} suggests that the AGB DTD of N
enrichment can safely be neglected for these purposes.
The~\no~abundance ratio in equilibrium is then given by the ratio of the yields
with the metallicity-dependent N yield evaluated at the equilibrium O abundance:
\begin{equation}
\no_\text{eq} = \log_{10}\left(\frac{
	Z_\text{N,eq} / Z_\text{O,eq}
}{
	Z_{\text{N},\odot} / {Z_{\text{O},\odot}}
}\right) = \log_{10}\left(\frac{
	y_\text{N}(Z_\text{O,eq}) / \ycc{O}
}{
	Z_{\text{N},\odot} / Z_{\text{O},\odot}
}\right),
\label{eq:noeq}
\end{equation}
where~$y_\text{N}(Z_\text{O})$ denotes the~\textit{total} IMF-averaged N yield
(CCSN + AGB) at a given~$Z_\text{O}$ as in Fig.~\ref{fig:ssp}.
We have spot-checked this formula for each of our AGB yield models taken from
the literature against the left panel of Fig.~\ref{fig:no_oh_predictions} and
found agreement to 0.02 dex or better, slightly smaller than the impact of
stellar migration on gas-phase N abundances (see discussion
in~\S\S~\ref{sec:results:fiducial} and~\ref{sec:results:schaefer_comp}).
\par
Equation~\refp{eq:noeq} can be used to predict the~\ohno~relation for a given
set of yields.
Our fiducial AGB star yield (equation~\ref{eq:linear_yield} with
$\xi = 9\times10^{-4}$) gives an IMF-averaged yield of~$9.3\times10^{-4}$ 
at~$Z = Z_\odot$ (Fig.~\ref{fig:ssp}; see also discussion
in~\S~\ref{sec:yields}).
Applying this value (along with~$\ycc{O} = 0.015$,
$Z_{\text{N},\odot} = 6.91\times10^{-4}$, and
$Z_{\text{O},\odot} = 5.72\times10^{-3}$) to equation~\refp{eq:noeq} and
separating the massive star contribution of~$\ycc{N} = 3.6\times10^{-4}$ for
generality gives
\begin{equation}
10^{\no_\text{eq}} = 10^{\no\subcc} + (0.513)10^{\oh_\text{eq}},
\end{equation}
where~$\no\subcc = -0.7$ is our empirical CCSN ``plateau'' taken from
Fig.~\ref{fig:no_oh_observed}, for which alternate values can be computed from
choices regarding~\ycc{N} and~\ycc{O} (see discussion
in~\S~\ref{sec:yields:ccsne}).
One can also reverse engineer equation~\refp{eq:noeq} to derive the IMF-averaged
N yield required to match an empirical~\ohno~relation given a value of~\ycc{O}.
\par
This analysis sharpens the conventional understanding of why~\no~is a useful
metallicity indicator in external galaxies when~\oh~cannot be measured directly.
The relation between~\no~and~\oh~is driven by stellar astrophysics (i.e. yields)
in a way that is relatively insensitive to the SFH, SFE, or other
galactic-scale physics.
Viewing the relation from this perspective also highlights where one should be
cautious about using~\no~as a metallicity indicator.
First is in environments where the stellar yields could be substantially
different, perhaps because of differences in the IMF or in stellar rotation.
Second is in galaxies that may be far from equilibrium, e.g. because of recent
bursts of star formation or mergers diluting the ISM.
The sinusoidal star formation variations considered
in~\S~\ref{sec:results:schaefer_comp} perturb equilibrium enough to create
$\sim$0.1 dex excursions in~\no~at fixed~\oh~(Fig.~\ref{fig:schaefer_comp}).
Third is in galaxies that are too young to have reached equilibrium.
Although the~\ohno~relation of our fiducial model is within~$\sim$0.1 dex of
its final value by~$t = 5$ Gyr, it is lower at earlier times
(Fig.~\ref{fig:no_oh_timeevol}).
One-zone models with a metallicity-dependent N yield can be used to estimate
whether high-redshift galaxies are likely to have reached equilibrium based
on their SFRs, gas fractions, metallicities, and stellar masses.
Dwarf galaxies often have low SFE and might therefore reach equilibrium more
slowly.
However, these galaxies are also typically in the low-metallicity regime
where~\no~is determined by the metallicity-independent primary yields anyway.

\section{Conclusions}
\label{sec:conclusions}

Building on the multi-zone GCE model of~\citet{Johnson2021}, which reproduces
many observed features of the [$\alpha$/Fe]-[Fe/H]-age distribution of Milky
Way disc stars, we have inferred empirical constraints on the stellar
nucleosynthesis of N by comparing model predictions to observed gas-phase trends
in extenal galaxies and stellar trends in the Milky Way disc.
In our models, the gas-phase abundance at a given Galactocentric radius first
evolves to higher~\oh~at roughly constant~\no~because of primary
(metallicity-independent) N production, then evolves upward in~\no~with slowly
increasing~\oh~because of secondary N production that increases with
metallicity (Fig.~\ref{fig:no_oh_timeevol}).
The~\ohno~relation reaches an approximate equilibrium after~$t = 5 - 8$ Gyr,
consistent with previous arguments that this relation is largely
redshift-independent~\citep{Vincenzo2018, HaydenPawson2021}.
This~\ohno~relation represents a superposition of evolutionary track endpoints
rather than an evolutionary track itself, similar to some explanations of the
low-$\alpha$ disc sequence in the Milky Way~\citep[e.g.][]{Schoenrich2009,
Nidever2014, Buck2020, Sharma2021, Johnson2021}.
\par
As our principal observational benchmark, we take Dopita et
al.'s (\citeyear{Dopita2016}) characterization of observed gas-phase abundances
in external galaxies (see Fig.~\ref{fig:no_oh_observed}).
Using Johnson et al.'s (\citeyear{Johnson2021}) CCSN oxygen yield
of~$\ycc{O} = 0.015$, we obtain agreement with
the~\citet{Dopita2016}~\ohno~relation if we assume a metallicity-independent
massive star yield of~$\ycc{N} = 3.6\times10^{-4}$ and an AGB fractional N
yield that is linear in stellar mass and metallicity
(equation~\ref{eq:linear_yield}) with~$\xi = 9\times10^{-4}$.
This value of~$\ycc{N}$ is consistent with the rotating massive star models of
\citet{Limongi2018}, and we concur with previous arguments that rotating massive
stars are required to explain the~$\no \approx -0.7$ plateau observed at low
metallicities (see Fig.~\ref{fig:no_oh_observed};~\citealp{Chiappini2003,
Chiappini2005, Chiappini2006, Kobayashi2011, Prantzos2018, Grisoni2021}).
The AGB yield is similar in form but 3 times higher in amplitude than the
models of~\cristallo.
\par
With~$\ycc{O} = 0.015$ and~$\ycc{N} = 3.6\times10^{-4}$, the AGB N yields
of~\cristallo~and~\ventura~must be amplified by factors of three and two,
respectively, to achieve agreement with the~\citet{Dopita2016}~\ohno~relation
(Fig.~\ref{fig:no_oh_predictions}).
However, as predicted abundance ratios depend primarily on yield ratios, we can
also obtain agreement by using the~\cristallo~or~\ventura~yields and
lowering~\ycc{O}~and~\ycc{N}~by the corresponding factor.
Such a change could be physically justified if black hole formation is more
extensive, or the IMF steeper, than implicitly assumed by the value
of~$\ycc{O} = 0.015$ (see~\S~\ref{sec:results:yields:variations}
and~\citealp{Griffith2021a}).
Other successful predictions of the~\citet{Johnson2021} models, including the
Galactic~\oh~gradient that is one of its basic constraints, would be largely
unchanged if~\ycc{Fe},~\yia{Fe}, and outflow mass loading efficiencies~$\eta$
were all reduced by the same factor.
Alternatively, one could retain a higher~\ycc{O}~and~\ycc{N}~but assume that
Galactic winds preferentially eject CCSN products relative to AGB products, as
suggested by~\citet{Vincenzo2016a}.
The degeneracy between the overall scaling of yields and the magnitude of
outflows is one of the key sources of uncertainty in GCE models.
\par
In contrast to~\cristallo~and~\ventura, the AGB models
of~\karakasten~and~\karakas~predict IMF-averaged yields that are decreasing or
approximately flat with increasing~$Z$ (Fig.~\ref{fig:ssp}).
In our GCE models, these yields lead to clear disagreement with the
\citet{Dopita2016} trend, even when we allow reasonable variations in the
metallicity dependence of~\ycc{N} (Fig.~\ref{fig:no_oh_predictions}).
There are many uncertain physical effects in AGB stellar models, so it is
difficult to pinpoint a single cause for this discrepancy.
In general, the most efficient N production occurs when both TDU and HBB occur
simultaneously because each replenishment of C and O isotopes from the stellar
core by TDU adds new seed nuclei for HBB to process in~\Nfourteen~via the CNO
cycle~\citep{Ventura2013}.
The distinctive metallicity dependence of the~\karakasten~and~\karakas~yields
traces back to the simultaneous occurrence of TDU and HBB over a substantial
mass range at all metallicities (Fig.~\ref{fig:agb_yield_models}).
\par
All of the AGB models predict that IMF-averaged N production is dominated by
stars with M~$> 2~\msun$ (Fig.~\ref{fig:ssp}).
As a result, the delay-time required to produce 50\% of the AGB N is 250 Myr or
less, shorter than the~$\sim$1 Gyr characteristic delay of Fe fron SN Ia.
The form of the~\ohno~relation is driven by the metallicity dependence of N
yields, not by the time delay of AGB production (Fig.~\ref{fig:t_z_dep_comp}),
and it can be calculated accurately from simple equilibrium arguments under
most circumstances (\S~\ref{sec:results:ohno_equilibrium},
equation~\ref{eq:noeq}).
\par
\citet{Vincenzo2021} inferred the median~\ohno~trend of Milky Way disc stars
from APOGEE abundances corrected for mixing on the red giant branch using the
asteroseismic mass measurements from~\citet{Miglio2021}.
They found good agreement with the~\citet{Dopita2016} trend, our observational
benchmark, so our model is also consistent with their derived APOGEE trends.
Our model also reproduces, at least approximately, two important findings of
\citet{Vincenzo2021}:~\no~exhibits little correlation with stellar age at
fixed~\feh~for ages~$\lesssim 9$ Gyr, and~\no~declines linearly with
increasing~\ofe~at fixed~\oh~(Fig.~\ref{fig:vincenzo_comp}).
The match to these observations~\textit{does} depend on the AGB DTD, and it
breaks down if we either make the AGB enrichment instantaneous or make it occur
as slowly as SN Ia Fe production.
\par
To investigate the sources of scatter in the~\ohno~relation, we construct
variants of our fiducial model that have~$\sim$40\% sinusoidal oscillations in
the SFR with a 2 Gyr period, induced by oscillations in either the SFE or the
gas infall rate.
The combined effects of dilution by pristine infall and metallicity-dependent N
production lead to oscillations in the~\ohno~relation comparable in magnitude
to the scatter measured in MaNGA galaxies by~\citet{Schaefer2020}
(Fig.~\ref{fig:schaefer_comp}).
We concur with their conclusion that variations in the SFE can plausibly
explain most of the observed scatter.
\citet{Johnson2021} find that stellar migration induces stochastic variations
in [$\alpha$/Fe] enrichment because a stellar population can migrate from its
birth radius before most of its SN Ia Fe production takes place.
The same effect occurs for AGB N enrichment but to a lesser extent because the
shorter production timescale ($\sim$250 Myr) leaves less time for migration.
We find that migration leads to~$\sim$0.05-dex scatter in~\no~at fixed~\oh,
which is smaller than the scatter measured by~\citet{Schaefer2020} but not
negligible.
\par
Our findings illustrate the value and methodology of empirically constraining
stellar yields by combining general theoretical expectations with GCE modelling
and observational constraints.
For the case of N, we have used the expectation that massive stars and AGB stars
both contribute, with the AGB contribution moderately delayed in time.
The metallicity dependence of the combined IMF-averaged yield is tightly
constrained, and it is plausibly partitioned into a massive star yield that is
independent of metallicity and an AGB yield that is linear in metallicity~$Z$
and progenitor mass~$M$.
The normalization of the yield is well-constrained~\textit{relative} to the
IMF-averaged O yield.
The DTD predicted by our fiducial model, in concert with the~\citet{Johnson2021}
GCE prescriptions, leads to good agreement with the~\no-age and~\no-\ofe~trends
for Milky Way disc stars (Fig.~\ref{fig:vincenzo_comp}).
As this approach is extended to increasing numbers of elements, the web of
yield constraints and consistency tests will become steadily more powerful,
providing valuable insights on stellar astrophysics, SN physics, and the
history of our Galaxy.

\section{Acknowledgements}
\label{sec:acknowledgements}

We are grateful to Amanda Karakas for valuable discussion on the physical
processes affecting N production in asymptotic giant branch stars.
We thank Paolo Ventura for providing theoretically predicted yields from
asymptotic giant branch stars at a wide variety of progenitor metallicities,
including unpublished tables at two of these metallicities.
We thank Brett Andrews for discussions of the MaNGA results that helped inspire
this study and for useful comments on the draft manuscript.
We also thank Adam Schaefer for providing us a copy of the data from
\citet{Schaefer2020}.
We acknowledge valuable discussion with Jennifer Johnson, Adam Leroy, Grace
Olivier, Amy Sardone, Jiayi Sun, Todd Thompson, and other members of The Ohio
State Astronomy Gas, Galaxies, and Feedback group.
This work was supported by National Science Foundation grant AST-1909841.
D.H.W. is grateful for the hospitality of the Institute for Advanced Study and
the support of the W.M. Keck Foundation and the Hendricks Foundation during
part of this work.
F.V. acknowledges the support of a Fellowship from the Center for Cosmology and
Astroparticle Physics at The Ohio State University.

\section{Data Availability}
\vice~is open-source software.
The code which numerically integrates the~\citet{Johnson2021} GCE models is
also publicly available and can be found in~\vice's GitHub repository.
The sample of star particles from~\hsim~is available through~\vice, which will
download the files automatically the first time it needs them.
The rest of the data from~\hsim~can be accessed
at~\url{https://nbody.shop/data.html}.
Observational data taken from the literature~\citep[e.g.][]{Dopita2016,
Schaefer2020, Vincenzo2021} will be provided upon request to the corresponding
author.

\bibliographystyle{mnras}
\bibliography{ms}

\begin{appendices}

\section{\vice}
\label{sec:vice}

\vice\footnote{
	Install (PyPI):~\url{https://pypi.org/project/vice} \\
	Documentation:~\url{https://vice-astro.readthedocs.io} \\
	Source Code:~\url{https://github.com/giganano/VICE.git}
} is an open-source~\python~package designed to model chemical enrichment
processes in galaxies with a generic, flexible model.
With this paper, we mark the release of version 1.3.0, which presents a handful
of new features:
\begin{enumerate}
	\item Users may select a mass-lifetime relation for stars from a list of
	several parametrized forms taken from the literature.
	Previously, only a single power-law was implemented, but this formulation
	underestimates lifetimes for stars with masses~$\gtrsim 4 M_\odot$; now,
	the options include the equations presented in:
	\begin{itemize}
		\item \citet{Vincenzo2016b}

		\item \citet*{Hurley2000}

		\item \citet{Kodama1997}

		\item \citet{Padovani1993}

		\item \citet{Maeder1989}

		\item \citet{Larson1974}~\textbf{(default)}
	\end{itemize}
	Generally, chemical evolution models make similar predictions with each of
	these different forms of the mass-lifetime relation since they are not
	considerably different from one another (see the section titled ``Single
	Stellar Populations'' under~\vice's science documentation for further
	discussion\footnote{\url
	{https://vice-astro.readthedocs.io/en/latest/science_documentation/}
	}).
	We select the~\citet{Larson1974} form as a default within~\vice~because it
	is representative of other forms and requires the lowest amount of
	computational overhead (aside from the single power-law option).
	This model is a metallicity-independent parabola in~$\log\tau - \log m$
	space for which we take updated coefficients from~\citet{Kobayashi2004}
	and~\citet{David1990}.

	\item We have added two additional tables of AGB star yields sampled at
	various progenitor masses and metallicities: 
	the~\karakas~and~\ventura~models presented in this paper are new
	to~\vice~(see discussion in~\S~\ref{sec:yields:agb} for details).

	\item We have built in the SN Ia yields presented in~\citet{Gronow2021a,
	Gronow2021b}.
	These tables include yields for double detonations of sub-Chandrasekhar
	mass carbon-oxygen white dwarfs at various progenitor metallicities.
\end{enumerate}
Although~\vice~includes built-in SN and AGB star yield tables, users are not
required to adopt any one of them for use in their chemical evolution models.
Instead, it allows arbitrary functions of metallicity for both CCSN and SN Ia
yields and functions of progenitor mass and metallicity for AGB star yields.
It provides similar flexibility for additional parameters typically built into
GCE models.
\vice's backend is implemented entirely in ANSI/ISO~\texttt{C}, providing it
with the powerful computing speeds of a compiled library while retaining such
scientific flexibility within the easy-to-use framework of~\python.
\par
Requiring a Unix kernel,~\vice~supports Mac and Linux operating systems;
Windows users should install and use~\vice~entirely within the Windows
Subsystem for Linux.
It can be installed via~\texttt{python -m pip install vice}, after which
\texttt{python -m vice -{}-docs} and~\texttt{python -m vice -{}-tutorial}
will launch a web browser to the documentation and to a jupyter notebook
intended to familiarize first time users with~\vice's API.

\end{appendices}

\label{lastpage}
\end{document}